\documentclass[twocolumn,showpacs,preprintnumbers,amsmath,amssymb,ctexart]{revtex4}
\usepackage{tipa}
\usepackage{graphicx}
\usepackage{graphics}
\usepackage{subfigure}
\usepackage{dcolumn}
\usepackage{slashbox}
\usepackage{bm}
\usepackage{amsmath}
\usepackage{color}
\usepackage{float}

\begin{document}

\title{Study of $2S$- and $1D$- excitations of observed charmed strange baryons}
\author{Dan-Dan Ye}
\affiliation{Department of Physics, Shanghai University, Shanghai 200444, China}
\affiliation{College of Mathematics, Physics and Information Engineering, Jiaxing University, Jiaxing 314001, China}

\author{Ze Zhao}
\affiliation{Department of Physics, Shanghai University, Shanghai 200444, China}


\author{Ailin Zhang}
\email{zhangal@staff.shu.edu.cn}
\affiliation{Department of Physics, Shanghai University, Shanghai 200444, China}

\begin{abstract}
Strong decays of $\Xi_c$ baryons with radial or orbital $\lambda$- and $\rho$- modes excitations with positive parity have been studied in a $^3P_0$ model.
As candidates of these kinds of excited charmed strange baryons, possible configurations and $J^P$ quantum numbers of $\Xi_c(2930)$, $\Xi_c(2980)$, $\Xi_c(3055)$, $\Xi_c(3080)$ and $\Xi_c(3123)$ have been assigned based on experimental data. There are $40$ kinds of configurations to describe the first radial or orbital excited $\Xi_c$ in $\lambda$- and $\rho$- mode excitations with positive parity. In these assignments, $\Xi_c(2930)$ may be a $2S$-wave excited $\tilde{\Xi}_{c1}(\frac{1}{2}^+)$ or $\tilde{\Xi}_{c1}(\frac{3}{2}^+)$, or a $D$-wave excited $\hat\Xi_{c1}^{' }(\frac{1}{2}^+)$, $\check\Xi_{c1}^{\ 0}(\frac{1}{2}^+)$, $\check\Xi_{c1}^{\ 2}(\frac{1}{2}^+)$, $\hat\Xi_{c1}^{' }(\frac{3}{2}^+)$, $\check\Xi_{c1}^{\ 0}(\frac{3}{2}^+)$ or $\check\Xi_{c1}^{\ 2}(\frac{3}{2}^+)$. $\Xi_c(2980)^+$ may be a $2S$-wave excited $\tilde{\Xi}_{c1}(\frac{1}{2}^+)$ or $\tilde{\Xi}_{c0}^{'}(\frac{1}{2}^+)$ with $J^P={1\over 2}^+$, or a $D$-wave excited $\check\Xi_{c0}^{'0}(\frac{1}{2}^+)$ or $\check\Xi_{c1}^{\ 0}(\frac{1}{2}^+)$ with $J^P={1\over 2}^+$. $\Xi_c(3055)^+$ may be a $2S$-wave excited $\acute{\Xi}_{c1}^{'}(\frac{3}{2}^+)$ or $\acute{\Xi}_{c0}(\frac{1}{2}^+)$. It may be a $D$-wave excited $\Xi_{c1}^{' }(\frac{3}{2}^+)$, $\Xi_{c2}^{' }(\frac{5}{2}^+)$, $\Xi_{c2}^{  }(\frac{3}{2}^+)$ or $\Xi_{c2}^{  }(\frac{5}{2}^+)$. $\Xi_c(3080)^+$ is very possibly a $2S$-wave excited $\acute{\Xi}_{c0}(\frac{1}{2}^+)$, and seems not a $D$-wave excitation of $\Xi_c$. For the poor experimental information of $\Xi_c(3123)$, it is impossible to identify this state at present. It is found that the channel $\Lambda D$ vanishes in the strong decay of $P$-wave, $D$-wave and $2S$-wave excited $\Xi_c$ without $\rho$- mode excitation between the two light quarks ($n_\rho=L_\rho=0$). In different configurations, some branching fraction ratios related to the internal structure of the $2S$-wave and $D$-wave of $\Xi_c$ are different. These ratios have been computed and can be employed to distinguish different configurations in forthcoming experiments.
\end{abstract}

\maketitle

\section{Introduction} \label{sec1}

In past years, many charmed strange baryons such as $\Xi_c(2790)$, $\Xi_c(2815)$, $\Xi_c(2980)$, $\Xi_c(3055)$ and $\Xi_c(3080)$ have been observed in different experiments~\cite{pdg}. Measurements of their masses, widths or some branching fraction ratios have been reported. Structure ($\Xi_c(2930)$) centered at an
invariant mass of $2.93$ GeV$/c^2$ in the $\Lambda^+_c K^-$ mass distribution of the decay $B^-\to \Lambda^+_c \bar\Lambda^-_c K^-$ was observed by BaBar collaboration~\cite{babar1}. Evidence for $\Xi_c(3123)$ decaying into intermediate-resonant mode $\Sigma_c(2520)K$ was found by BaBar collaboration~\cite{babar2} though no signature was seen in the same mode by Belle collaboration~\cite{belle3}. Unfortunately, neither the spin nor the parity of these $\Xi_c$ has been measured.

In these $\Xi_c$, $\Xi_c(2790)$ and $\Xi_c(2815)$ are believed the $P$-wave excited $\Xi_c$ with $J^P={1\over 2}^-$ and $J^P={3\over 2}^-$, respectively. However, there are different interpretations of $\Xi_c(2930)$, $\Xi_c(2980)$, $\Xi_c(3055)$, $\Xi_c(3080)$ and $\Xi_c(3123)$~\cite{brac,petry,ebert,roberts,guo,huang,zhang1,zhu1,zhang2,zhu2,shah,zhang3,cheng,zhu3,zhong,zhang4,zhu4,wang,ye1,zhong2}.
Study of their hadronic decays is an important way to understand the structure and dynamics of these excited $\Xi_c$.

$^3P_0$ model as a phenomenological method has been employed successfully to study the Okubo-Zweig-Iizuka (OZI)-allowed hadronic decays of hadrons~\cite{micu1969,yaouanc1,yaouanc2,yaouanc3}.
In Ref.~\cite{ye1}, $\Xi_c(2930)$, $\Xi_c(2980)$, $\Xi_c(3055)$, $\Xi_c(3080)$ and $\Xi_c(3123)$ as possible $P$-wave excited $\Xi_c$ candidates have been systematically studied through their strong decay in the model, but other possible assignments of these excited $\Xi_c$ are reserved. As a subsequent study, possible $2S$-wave or $D$-wave excitation assignments will be explored in this paper, which consists of a comprehensive understanding of the observed $\Xi_c(2930)$, $\Xi_c(2980)$, $\Xi_c(3055)$, $\Xi_c(3080)$ and $\Xi_c(3123)$.

The paper is organized as follows. In Sec. II, we give a brief review of the $^3P_0$ model and present some notations of $\Xi_c$ baryons. Numerical results for hadronic decays of these $\Xi_c$ are presented in Sec. III. In the last section, we give our conclusions and discussions.

\section{The $^3P_0$ model and some notations of $\Xi_c$ baryons}

Although the intrinsic mechanism and the relation to the Quantum Chromodynamics are not clear, the $^3P_0$ model has been employed successfully to study the OZI-allowed strong decays of hadrons by many authors after its proposal~\cite{micu1969, yaouanc1,yaouanc2,yaouanc3}.

There are three possible rearrangements for the decay process of a $\Xi_c$ baryon as indicated in Ref.~\cite{ye1},

\begin{eqnarray}\label{eq1}
A(q_1s_2c_3)+P(q_4\overline{q}_5)\to B(q_4s_2c_3)+C(q_1\overline{q}_5)\\
A(q_1s_2c_3)+P(q_4\overline{q}_5)\to B(q_1q_4c_3)+C(s_2\overline{q}_5) \\
A(q_1s_2c_3)+P(q_4\overline{q}_5)\to B(q_1q_4s_2)+C(c_3\overline{q}_5),\label{eq3}
\end{eqnarray}
where $q_i$ denotes a $u$ or $d$ quark. $s_2$ and $c_3$ denote a strange and a charm quark, respectively.

In the $^3P_0$ model, a hadronic decay width $\Gamma$ of a process $A \to B + C$ follows~\cite{yaouanc3},
\begin{eqnarray}
\Gamma  = \pi ^2 \frac{|\vec{p}|}{m_A^2} \frac{1}{2J_A+1}\sum_{M_{J_A}M_{J_B}M_{J_C}} |{\mathcal{M}^{M_{J_A}M_{J_B}M_{J_C}}}|^2
\end{eqnarray}
with a helicity amplitude $\mathcal{M}^{M_{J_A}M_{J_B}M_{J_C}}$~\cite{zhu3,zhang4,zhang5,ye1}
\begin{flalign}
 &\delta^3(\vec{p}_B+\vec{p}_C-\vec{p}_A)\mathcal{M}^{M_{J_A } M_{J_B } M_{J_C }}\nonumber \\
 &=-\gamma\sqrt {8E_A E_B E_C }  \sum_{M_{\rho_A}}\sum_{M_{L_A}}\sum_{M_{\rho_B}}\sum_{M_{L_B}} \sum_{M_{S_1},M_{S_3},M_{S_4},m}  \nonumber\\
 &\langle {J_{l_A} M_{J_{l_A} } S_3 M_{S_3 } }| {J_A M_{J_A } }\rangle \langle {L_{\rho_A} M_{L_{\rho_A} } L_{\lambda_A} M_{L_{\lambda_A} } }| {L_A M_{L_A } }\rangle \nonumber \\
 &\langle L_A M_{L_A } S_{12} M_{S_{12} }|J_{l_A} M_{J_{l_A} } \rangle \langle S_1 M_{S_1 } S_2 M_{S_2 }|S_{12} M_{S_{12} }\rangle \nonumber \\
 &\langle {J_{l_B} M_{J_{l_B} } S_3 M_{S_3 } }| {J_B M_{J_B } }\rangle \langle {L_{\rho_B} M_{L_{\rho_B} } L_{\lambda_B} M_{L_{\lambda_B} } }| {L_B M_{L_B } }\rangle \nonumber \\
 &\langle L_B M_{L_B } S_{14} M_{S_{14} }|J_{l_B} M_{J_{l_B}}  \rangle \langle S_1 M_{S_1 } S_4 M_{S_4 }|S_{14} M_{S_{14} }\rangle \nonumber \\
 &\langle {1m;1 - m}|{00} \rangle \langle S_4 M_{S_4 } S_5 M_{S_5 }|1 -m \rangle \nonumber \\
 &\langle L_C M_{L_C } S_C M_{S_C}|J_C M_{J_C} \rangle \langle S_2 M_{S_2 } S_5 M_{S_5 }|S_C M_{S_C} \rangle \nonumber \\
&\times\langle\varphi _B^{1,4,3} \varphi _C^{2,5}|\varphi _A^{1,2,3}\varphi _0^{4,5} \rangle \times I_{M_{L_B } ,M_{L_C } }^{M_{L_A },m} (\vec{p})
\end{flalign}
in which the spatial integral
\begin{flalign}
I_{M_{L_B } ,M_{L_C } }^{M_{L_A } ,m} (\vec{p})&= \int d \vec{p}_1 d \vec{p}_2 d \vec{p}_3 d \vec{p}_4 d \vec{p}_5 \nonumber \\
&\times\delta ^3 (\vec{p}_1 + \vec{p}_2 + \vec{p}_3 -\vec{p}_A)\delta ^3 (\vec{p}_4+ \vec{p}_5)\nonumber \\
&\times \delta ^3 (\vec{p}_1 + \vec{p}_4 + \vec{p}_3 -\vec{p}_B )\delta ^3 (\vec{p}_2 + \vec{p}_5 -\vec{p}_C) \nonumber \\
& \times\Psi _{B}^* (\vec{p}_1, \vec{p}_4,\vec{p}_3)\Psi _{C}^* (\vec{p}_2 ,\vec{p}_5) \nonumber \\
& \times \Psi _{A} (\vec{p}_1 ,\vec{p}_2 ,\vec{p}_3)y _{1m}\left(\frac{\vec{p_4}-\vec{p}_5}{2}\right).
\end{flalign}
Details of the indices, matrix elements, and other notations in the $^3P_0$ model can
be found in Refs.~\cite{yaouanc3,Capstick1993,zhu3,zhang4,zhang5,ye1}.

In our calculation, notations for the excited baryons are the same as those in Refs.~\cite{cheng,zhu3,zhang4,ye1}. In Table \ref{table1}, $n_\rho$ and $L_\rho$ denote the nodal and the orbital angular momentum between the two light quarks, $n_\lambda$ and $L_\lambda$ denote the nodal and the orbital angular momentum between the charm quark and the two light quark system. $L$ is the total orbital angular momentum of $L_\rho$ and $L_\lambda$, $S_\rho$ denotes the total spin of the two light quarks, and $J_l$ is total angular momentum of $L$ and $S_\rho$. $J$ is the total angular momentum of the baryons. In Table~\ref{table1}, the acute, the tilde, the hat and the check are employed to denote the baryons with $n_\lambda=1$, $n_\rho=1$, $L_\rho=2$ and $L_\rho=1$, respectively. For $\check\Xi_{cJ_l}^{\ L}$, a superscript $L$ is specialized to denote different total angular momentum. For $\Xi_c$ baryons, there are $40$ different configurations to describe the first radial ($6$ types, denoted as $2S$-wave in the following) or orbital ($34$ types, denoted as $D$-wave) excitations in $\lambda$- and $\rho$- modes with positive parity. The $J^P$ quantum numbers of these $2S$- and $D$-wave charmed strange baryons are also given in Table~\ref{table1}.

\begin{table}[htbp]
\caption{Quantum numbers of $2S$-wave and $1D$-wave excitations}
\begin{tabular}{p{0.0cm} p{2.5cm}*{8}{p{0.5cm}}}
   \hline\hline
             & Assignments                                         & $J$                         & $J_l$ & $n_\rho$ & $L_\rho$ & $n_\lambda$ & $L_\lambda$ & $L$  & $S_\rho$ \\

\hline
\label{35}   &$\acute{\Xi}_{c1}^{'}(\frac{1}{2}^+)$               & $\frac{1}{2}$   &1    &  0 & 0  & 1 &  0 &  1 &  1   \\

\label{36}   &$\acute{\Xi}_{c1}^{'}(\frac{3}{2}^+)$               &$\frac{3}{2}$    &1    &  0 & 0  & 1 &  0 &  1 &  1  \\

\label{37}   &$\acute{\Xi}_{c0}(\frac{1}{2}^+)$                   & $\frac{1}{2}$   &0    &  0 & 0  & 1 &  0 &  1 &  0   \\

\label{38}   &$\tilde{\Xi}_{c1}(\frac{1}{2}^+)$                   & $\frac{1}{2}$   &1    &  1 & 0  & 0 &  0 &  1 &  1   \\

\label{39}   &$\tilde{\Xi}_{c1}(\frac{3}{2}^+)$                   & $\frac{3}{2}$   &1    &  1 & 0  & 0 &  0 &  1 &  1  \\

\label{40}   &$\tilde{\Xi}_{c0}^{'}(\frac{1}{2}^+)$               & $\frac{1}{2}$   &0    &  1 & 0  & 0 &  0 &  1 &  0  \\

   \hline
   \hline
\label{01}   &$\Xi_{c1}^{' }(\frac{1}{2}^+,\frac{3}{2}^+)$        & $\frac{1}{2}$,$\frac{3}{2}$ &  1    &  0       &  0       &  0       &   2         &  2   &  1       \\
\label{03}   &$\Xi_{c2}^{' }(\frac{3}{2}^+,\frac{5}{2}^+)$        & $\frac{3}{2}$,$\frac{5}{2}$ &  2    &  0       &  0       &  0       &   2         &  2   &  1       \\
\label{05}   &$\Xi_{c3}^{' }(\frac{5}{2}^+,\frac{7}{2}^+)$        & $\frac{3}{2}$,$\frac{5}{2}$ &  3    &  0       &  0       &  0       &   2         &  2   &  1       \\
\label{07}   &$\Xi_{c2}^{  }(\frac{3}{2}^+,\frac{5}{2}^+)$        & $\frac{3}{2}$,$\frac{5}{2}$ &  2    &  0       &  0       &  0       &   2         &  2   &  0       \\
\label{09}   &$\hat\Xi_{c1}^{' }(\frac{1}{2}^+,\frac{3}{2}^+)$    & $\frac{1}{2}$,$\frac{3}{2}$ &  1    &  0       &  2       &  0       &   0         &  2   &  1       \\
\label{11}   &$\hat\Xi_{c2}^{' }(\frac{3}{2}^+,\frac{5}{2}^+)$    & $\frac{3}{2}$,$\frac{5}{2}$ &  2    &  0       &  2       &  0       &   0         &  2   &  1       \\
\label{13}   &$\hat\Xi_{c3}^{' }(\frac{5}{2}^+,\frac{7}{2}^+)$    & $\frac{3}{2}$,$\frac{5}{2}$ &  3    &  0       &  2       &  0       &   0         &  2   &  1       \\
\label{15}   &$\hat\Xi_{c2}^{ }(\frac{3}{2}^+,\frac{5}{2}^+)$     & $\frac{3}{2}$,$\frac{5}{2}$ &  2    &  0       &  2       &  0       &   0         &  2   &  0       \\
\label{17}   &$\check\Xi_{c0}^{'0}(\frac{1}{2}^+)$                & $\frac{1}{2}$               &  0    &  0       &  1       &  0       &   1         &  0   &  0       \\
\label{18}   &$\check\Xi_{c1}^{'1}(\frac{1}{2}^+,\frac{3}{2}^+)$  & $\frac{1}{2}$,$\frac{3}{2}$ &  1    &  0       &  1       &  0       &   1         &  1   &  0       \\
\label{20}   &$\check\Xi_{c2}^{'2}(\frac{3}{2}^+,\frac{5}{2}^+)$  & $\frac{3}{2}$,$\frac{5}{2}$ &  2    &  0       &  1       &  0       &   1         &  2   &  0       \\
\label{22}   &$\check\Xi_{c1}^{\ 0}(\frac{1}{2}^+,\frac{3}{2}^+)$ & $\frac{1}{2}$,$\frac{3}{2}$ &  1    &  0       &  1       &  0       &   1         &  0   &  1       \\
\label{24}   &$\check\Xi_{c0}^{\ 1}(\frac{1}{2}^+)$               & $\frac{1}{2}$               &  0    &  0       &  1       &  0       &   1         &  1   &  1       \\
\label{25}   &$\check\Xi_{c1}^{\ 1}(\frac{1}{2}^+,\frac{3}{2}^+)$ & $\frac{1}{2}$,$\frac{3}{2}$ &  1    &  0       &  1       &  0       &   1         &  1   &  1       \\
\label{27}   &$\check\Xi_{c2}^{\ 1}(\frac{3}{2}^+,\frac{5}{2}^+)$ & $\frac{3}{2}$,$\frac{5}{2}$ &  2    &  0       &  1       &  0       &   1         &  1   &  1       \\
\label{29}   &$\check\Xi_{c1}^{\ 2}(\frac{1}{2}^+,\frac{3}{2}^+)$ & $\frac{1}{2}$,$\frac{3}{2}$ &  1    &  0       &  1       &  0       &   1         &  2   &  1       \\
\label{31}   &$\check\Xi_{c2}^{\ 2}(\frac{3}{2}^+,\frac{5}{2}^+)$ & $\frac{3}{2}$,$\frac{5}{2}$ &  2    &  0       &  1       &  0       &   1         &  2   &  1       \\
\label{33}   &$\check\Xi_{c3}^{\ 2}(\frac{5}{2}^+,\frac{7}{2}^+)$ & $\frac{5}{2}$,$\frac{7}{2}$ &  3    &  0       &  1       &  0       &   1         &  2   &  1       \\
\hline\hline
\end{tabular}
\label{table1}
\end{table}

The parameters involved in the $^3P_0$ model are chosen the same as those in Ref.~\cite{ye1}, where the experimental data of $\Xi_c(2645)$, $\Xi_c(2790)$ and $\Xi_c(2815)$ is well reproduced. The input of masses of baryons and mesons in this work are from the PDG~\cite{pdg}.

\section{Hadronic decays of $\Xi_c$ baryons}

As candidates of $2S$ or $D$ excitations of $\Xi_c$, main decay modes and relevant hadronic decay widths of $\Xi_c(2930)$, $\Xi_c(2980)$, $\Xi_c(3055)^+$, $\Xi_c(3080)^+$ and $\Xi_c(3123)^+$ in different assignments are computed and presented in Tables~\ref{table2}-\ref{table11}, respectively.
The vanish modes in these tables indicate forbidden channels or channels with very small decay width. The ``$\cdots$" in these tables indicates that there exists no such term. $\Gamma_{sum}$ indicates summation of presented partial decay widths. Some ratios of branching fractions related to experiments are particularly given.

\subsubsection{ $\Xi_c(2930)$}

The signal of $\Xi_c(2930)$ was found in the $\Lambda_c^+ K^-$ mass distribution with $\Gamma=36\pm 7\pm 11$ MeV by BaBar collaboration~\cite{babar1}. So far, this signal has not been observed by any other experiment.

If $\Xi_c(2930)$ is a $2S$-wave excitation, its OZI-allowed hadronic decay channels and widths are presented in Table \ref{table2}. From this table, the $\tilde{\Xi}_{c0}^{'}(\frac{1}{2}^+)$ or $\acute{\Xi}_{c0}(\frac{1}{2}^+)$ possibility of $\Xi_c(2930)$ is excluded since the theoretically forbidden decay mode $\Lambda_c^+ K^-$ was observed by experiment. The $\acute{\Xi}_{c1}(\frac{1}{2}^+)$ or $\acute{\Xi}_{c1}(\frac{3}{2}^+)$ possibility is excluded either for a narrow decay width. Our results indicate that $\Xi_c(2930)$ may be a $\tilde{\Xi}_{c1}(\frac{1}{2}^+)$ or $\tilde{\Xi}_{c1}(\frac{3}{2}^+)$. In order to distinguish these two possibilities, measurements of the different ratios, $\mathcal{B}_2=B(\Xi_{c}(2930)^{+} \to \Xi_c^{'0} \pi^{+})/B(\Xi_{c}(2930)^{+} \to \Xi_{c}(2645)^{0}\pi^{+})=2.5\sim 4.1$ for $\tilde{\Xi}_{c1}(\frac{1}{2}^+)$ and the $\mathcal{B}_2=0.3\sim 0.4$ for $\tilde{\Xi}_{c1}(\frac{3}{2}^+)$, will be helpful.

\begin{center}
  \begin{table*}
     \caption{Decay widths (MeV) of $ \Xi_c(2930)^+$ as the $2S$-wave excitations.  $\mathcal{B}_2=B(\Xi_{c}(2930)^{+} \to \Xi_c^{'0} \pi^{+})/B(\Xi_{c}(2930)^{+} \to \Xi_{c}(2645)^{0}\pi^{+})$}
     \begin{tabular}{c |ccccc|cccccccccccccccccc} \hline \hline
 $ \Xi_{cJ_l} (J^P) $ & $ n_{\lambda} $ & $L_\lambda$  & $ n_{\rho} $ & $L_\rho$   & $S_\rho$
 &$\Xi_c\pi    $ &$\Xi_c^{'}\pi $ &$\Xi_c(2645)\pi$  &$\Lambda_c K$    &$\Gamma_{total}$   & $\mathcal{B}_2 $  \\
     \hline
  $\acute{\Xi}_{c1}^{'}(\frac{1}{2}^+)$  & 1 &  0 &  0 & 0  &  1  &$0.1\sim2.9$  &$0.4\sim2.1$  &$0.2\sim0.5$    &$0.3\sim2.3$     &$1.0\sim7.8$        &$2.5\sim4.1$ \\
  $\acute{\Xi}_{c1}^{'}(\frac{3}{2}^+)$  & 1 &  0 &  0 & 0  &  1  &$0.1\sim2.9$  &$0.1\sim0.5$  &$0.5\sim1.3$    &$0.3\sim2.3$     &$1.0\sim7.0$        &$0.3\sim0.4$\\
  $\acute{\Xi}_{c0}(\frac{1}{2}^+)$      & 1 &  0 &  0 & 0  &  0  &0             &$0.3\sim1.6$  &$0.6\sim1.5$               &0                &$0.9\sim3.1$        &$0.6\sim1.2$ \\

  $\tilde{\Xi}_{c1}(\frac{1}{2}^+)$& 0 &0 &  1 & 0  &  1          &$1.2\sim26$    &$4.2\sim19$   &$1.7\sim4.6$    &$2.3\sim21$      &$9.3\sim70$         &$2.5\sim4.1$ \\
  $\tilde{\Xi}_{c1}(\frac{3}{2}^+)$& 0 &0 &  1 & 0  &  1          &$1.2\sim26$    &$1.0\sim4.7$  &$4.2\sim11$     &$2.3\sim21$      &$8.7\sim63$         &$0.3\sim0.4$\\
  $\tilde{\Xi}_{c0}^{'}(\frac{1}{2}^+)$& 0&0&1 & 0  &  0          &0              &$3.1\sim14$   &$5.1\sim14$               &0                &$8.2\sim28$         &$0.6\sim1.0$\\

    \hline\hline
    \end{tabular}
    \label{table2}
  \end{table*}
\end{center}

If $\Xi_c(2930)$ is a $D$-wave excitation, its OZI-allowed hadronic decay channels and widths are presented in Table \ref{table3}. Obviously, the $20$ possibilities with a theoretically forbidden $\Lambda_c^+ K^-$ decay channel which was observed by experiment seem impossible. The $6$ configurations with $j_l=3$ have tiny total decay widths, so the assignments of $\Xi_c(2930)$ with $\Xi_{c3}^{' }(\frac{5}{2}^+)$, $\Xi_{c3}^{' }(\frac{7}{2}^+)$, $\hat\Xi_{c3}^{' }(\frac{5}{2}^+)$, $\hat\Xi_{c3}^{' }(\frac{7}{2}^+)$, $\check\Xi_{c3}^{\ 2}(\frac{5}{2}^+)$ or $\check\Xi_{c3}^{\ 2}(\frac{7}{2}^+)$ could be excluded. The $\Xi_{c1}^{' }(\frac{1}{2}^+)$ or $\Xi_{c1}^{' }(\frac{3}{2}^+)$ possibility seems impossible either for a narrow width. There are $6$ possible assignments left for $\Xi_c(2930)$. In other words, $\Xi_c(2930)$ may be a $\hat\Xi_{c1}^{' }(\frac{1}{2}^+)$, $\check\Xi_{c1}^{\ 0}(\frac{1}{2}^+)$ or $\check\Xi_{c1}^{\ 2}(\frac{1}{2}^+)$ with quantum numbers $J^P={1\over 2}^+$, it could be a $\hat\Xi_{c1}^{' }(\frac{3}{2}^+)$, $\check\Xi_{c1}^{\ 0}(\frac{3}{2}^+)$ or $\check\Xi_{c1}^{\ 2}(\frac{3}{2}^+)$ with quantum numbers $J^P={3\over 2}^+$. With these two different $J^P$ quantum numbers, the ratios $\mathcal{B}_2=B(\Xi_{c}(2930)^{+} \to \Xi_c^{'0} \pi^{+})/B(\Xi_{c}(2930)^{+} \to \Xi_{c}(2645)^{0}\pi^{+})$ are different. Therefore, measurements of these ratios in the future are important for the identification of $\Xi_c(2930)$.

\begin{center}
  \begin{table*}\footnotesize
     \caption{Decay widths (MeV) of $ \Xi_c(2930)^+$ as the $D$-wave excitations.
     $\mathcal{B}_2=B(\Xi_{c}(2930)^{+} \to \Xi_c^{'0} \pi^{+})/B(\Xi_{c}(2930)^{+} \to \Xi_{c}(2645)^{0}\pi^{+})$  }
     \begin{tabular}{c |ccccc|cccccccccccccccccc} \hline \hline
 $ \Xi_{cJ_l} (J^P) $ & $ n_{\lambda} $ & $L_\lambda$  & $ n_{\rho} $ & $L_\rho$   & $S_\rho$
 &$\Xi_c\pi    $ &$\Xi_c^{'}\pi $ &$\Xi_c(2645)\pi$  &$\Lambda_c K$    &$\Gamma_{total}$   & $\mathcal{B}_2 $  \\
     \hline

\label{01}$\Xi_{c1}^{' }(\frac{1}{2}^+)$  & 0 &  2 &  0 & 0  &  1      &$2.0\sim5.2$   &$0.6\sim0.8$    &$0.2\sim0.2$        &$1.9\sim3.9$     &$4.7\sim10$   &$3.3\sim4.3$\\  
\label{02}$\Xi_{c1}^{' }(\frac{3}{2}^+)$  & 0 &  2 &  0 & 0  &  1      &$2.0\sim5.2$   &$0.1\sim0.2$    &$0.4\sim0.5$        &$1.9\sim3.9$     &$4.5\sim9.9$  &$0.3\sim0.4$\\  
\label{03}$\Xi_{c2}^{' }(\frac{3}{2}^+)$  & 0 &  2 &  0 & 0  &  1      &0              &$1.3\sim1.9$    &$0.2\sim0.2$        &0                &$1.4\sim2.1$  &$7.6\sim10$\\  
\label{04}$\Xi_{c2}^{' }(\frac{5}{2}^+)$  & 0 &  2 &  0 & 0  &  1      &0              &$0.05\sim0.01$  &$0.9\sim1.1$        &0                &$1.0\sim1.1$  &$0.05\sim0.01$\\  
\label{05}$\Xi_{c3}^{' }(\frac{5}{2}^+)$  & 0 &  2 &  0 & 0  &  1      &$0.6\sim0.2$   &$0.06\sim0.01$  &$0.01\sim0.0$       &$0.3\sim0.08$    &$0.9\sim0.3$  &$4.5\sim5.1$\\  
\label{06}$\Xi_{c3}^{' }(\frac{7}{2}^+)$  & 0 &  2 &  0 & 0  &  1      &$0.6\sim0.2$   &$0.03\sim0.0$   &$0.02\sim0.0$      &$0.3\sim0.08$    &$0.9\sim0.3$  &$1.9\sim2.1$\\  
\label{07}$\Xi_{c2}^{  }(\frac{3}{2}^+)$  & 0 &  2 &  0 & 0  &  0      &0              &$0.8\sim1.3$    &$0.1\sim0.1$        &0                &$1.0\sim1.4$  &$6.8\sim10$\\  
\label{08}$\Xi_{c2}^{  }(\frac{5}{2}^+)$  & 0 &  2 &  0 & 0  &  0      &0              &$0.07\sim0.02$  &$0.6\sim0.7$        &0                &$0.7\sim0.7$  &$0.1\sim0.02$\\  
\label{09}$\hat\Xi_{c1}^{' }(\frac{1}{2}^+)$ & 0 &  0 &  0 & 2  &  1   &$18\sim47$     &$5.1\sim7.6$    &$1.5\sim1.8$        &$17\sim35$       &$42\sim91$    &$3.3\sim4.3$\\  
\label{10}$\hat\Xi_{c1}^{' }(\frac{3}{2}^+)$ & 0 &  0 &  0 & 2  &  1   &$18\sim47$     &$1.3\sim1.9$    &$3.8\sim4.5$        &$17\sim35$       &$40\sim88$    &$0.3\sim0.4$\\  
\label{11}$\hat\Xi_{c2}^{' }(\frac{3}{2}^+)$ & 0 &  0 &  0 & 2  &  1   &0              &$11\sim17$      &$1.5\sim1.7$        &0                &$13\sim19$    &$7.6\sim10$  \\  
\label{12}$\hat\Xi_{c2}^{' }(\frac{5}{2}^+)$ & 0 &  0 &  0 & 2  &  1   &0              &$0.4\sim0.1$    &$8.3\sim9.7$          &0                &$8.8\sim9.8$  &$0.05\sim0.01$\\  
\label{13}$\hat\Xi_{c3}^{' }(\frac{5}{2}^+)$ & 0 &  0 &  0 & 2  &  1   &$5.3\sim1.7$   &$0.5\sim0.1$    &$0.1\sim0.03$       &$2.6\sim0.8$     &$8.6\sim2.6$  &$4.6\sim5.1$\\  
\label{14}$\hat\Xi_{c3}^{' }(\frac{7}{2}^+)$ & 0 &  0 &  0 & 2  &  1   &$5.3\sim1.7$   &$0.3\sim0.07$   &$0.2\sim0.03$        &$2.6\sim0.8$     &$8.4\sim2.6$  &$1.9\sim2.1$\\  
\label{15}$\hat\Xi_{c2}^{ }(\frac{3}{2}^+)$  & 0 &  0 &  0 & 2  &  0   &0              &$7.6\sim11$     &$1.1\sim1.1$        &0                &$8.7\sim13$   &$6.8\sim10$ \\  
\label{16}$\hat\Xi_{c2}^{ }(\frac{5}{2}^+)$  & 0 &  0 &  0 & 2  &  0   &0              &$0.7\sim0.2$    &$5.6\sim6.5$         &0                &$6.3\sim6.7$  &$0.1\sim0.03$\\  
\label{17}$\check\Xi_{c0}^{'0}(\frac{1}{2}^+)$ & 0 &  1 &  0 & 1  &  0 &0              &$2.1\sim9.4$    &$3.4\sim9.2$         &0                &$5.5\sim19$   &$0.6\sim1.0$\\  
\label{18}$\check\Xi_{c1}^{'1}(\frac{1}{2}^+)$ & 0 &  1 &  0 & 1  &  0 &0              &0               &0                  &0              &0       &0    \\  
\label{19}$\check\Xi_{c1}^{'1}(\frac{3}{2}^+)$ & 0 &  1 &  0 & 1  &  0 &0              &0               &0                  &0              &0       &0    \\  
\label{20}$\check\Xi_{c2}^{'2}(\frac{3}{2}^+)$ & 0 &  1 &  0 & 1  &  0 &0              &$5.1\sim7.6$    &$0.8\sim0.8$        &0                &$5.8\sim8.4$  &$6.8\sim10$\\  
\label{21}$\check\Xi_{c2}^{'2}(\frac{5}{2}^+)$ & 0 &  1 &  0 & 1  &  0 &0              &$0.4\sim0.1$    &$3.7\sim4.3$        &0                &$4.2\sim4.4$  &$0.1\sim0.03$\\  
\label{22}$\check\Xi_{c1}^{\ 0}(\frac{1}{2}^+)$& 0 &  1 &  0 & 1  &  1 &$0.8\sim17$    &$2.8\sim12$     &$1.1\sim3.1$        &$1.5\sim14$      &$6.2\sim47$   &$2.5\sim4.1$\\  
\label{23}$\check\Xi_{c1}^{\ 0}(\frac{3}{2}^+)$& 0 &  1 &  0 & 1  &  1 &$0.8\sim17$    &$0.7\sim3.1$    &$2.8\sim7.8$         &$1.5\sim14$      &$5.8\sim42$   &$0.3\sim0.4$\\  
\label{24}$\check\Xi_{c0}^{\ 1}(\frac{1}{2}^+)$ & 0 &  1 &  0 & 1  &  1&0              &0               &0          &0      &0     &0    \\  
\label{25}$\check\Xi_{c1}^{\ 1}(\frac{1}{2}^+)$ & 0 &  1 &  0 & 1  &  1&0              &0               &0          &0      &0     &0    \\  
\label{26}$\check\Xi_{c1}^{\ 1}(\frac{3}{2}^+)$ & 0 &  1 &  0 & 1  &  1&0              &0               &0          &0      &0     &0    \\  
\label{27}$\check\Xi_{c2}^{\ 1}(\frac{3}{2}^+)$ & 0 &  1 &  0 & 1  &  1&0              &0               &0          &0      &0     &0    \\  
\label{28}$\check\Xi_{c2}^{\ 1}(\frac{5}{2}^+)$ & 0 &  1 &  0 & 1  &  1&0              &0               &0          &0      &0     &0    \\  
\label{29}$\check\Xi_{c1}^{\ 2}(\frac{1}{2}^+)$ & 0 &  1 &  0 & 1  &  1&$12\sim31$     &$3.4\sim5.1$    &$1.0\sim1.2$         &$12\sim23$      &$28\sim61$     &$3.3\sim4.3$\\  
\label{30}$\check\Xi_{c1}^{\ 2}(\frac{3}{2}^+)$ & 0 &  1 &  0 & 1  &  1&$12\sim31$     &$0.8\sim1.3$    &$2.5\sim3.0$         &$12\sim23$      &$27\sim59$     &$0.3\sim0.4$\\  
\label{31}$\check\Xi_{c2}^{\ 2}(\frac{3}{2}^+)$ & 0 &  1 &  0 & 1  &  1&0              &$7.6\sim11$     &$1.0\sim1.1$         &0               &$8.6\sim13$    &$7.6\sim10$ \\  
\label{32}$\check\Xi_{c2}^{\ 2}(\frac{5}{2}^+)$ & 0 &  1 &  0 & 1  &  1&0              &$0.3\sim0.08$   &$5.5\sim6.5$          &0               &$5.8\sim6.6$   &$0.05\sim0.01$\\  
\label{33}$\check\Xi_{c3}^{\ 2}(\frac{5}{2}^+)$ & 0 &  1 &  0 & 1  &  1&$3.6\sim1.1$   &$0.3\sim0.09$   &$0.08\sim0.02$        &$1.7\sim0.5$    &$5.7\sim1.7$   &$4.6\sim5.1$\\  
\label{34}$\check\Xi_{c3}^{\ 2}(\frac{7}{2}^+)$ & 0 &  1 &  0 & 1  &  1&$3.6\sim1.1$   &$0.2\sim0.05$   &$0.1\sim0.02$        &$1.7\sim0.5$    &$5.6\sim1.7$   &$1.9\sim2.1$\\  
    \hline\hline
    \end{tabular}
    \label{table3}
  \end{table*}
\end{center}

In Ref.~\cite{ye1}, the assignments of $\Xi_c(2930)$ with the $P$-wave excitations ($J^P={1\over 2}^-$, ${3\over 2}^-$ and ${3\over 2}^-$) were also advocated in the $^3P_0$ model. Therefore, $\Xi_c(2930)$ may be a $P$-wave, $D$-wave, or $2S$-wave excitation of $\Xi_c$. More experimental information on the branching fraction ratios is required for its identification.

\subsubsection{ $\Xi_c(2980)$}

$\Xi_c(2980)^+$ was first observed by Belle collaboration in the $\Lambda_c^+ K^- \pi^+ $ channel with $\Gamma=43.5\pm 7.5\pm 7.0$ MeV~\cite{Chistov:0606051} and confirmed by BaBar collaboration in the intermediate resonant mode $\Sigma_c(2455)^+K^-$~\cite{babar1}. It was subsequently observed by Belle collaboration in the $\Xi_c(2645)^0\pi^+$ decay channel~\cite{Belle:0802.3968}. In Ref.~\cite{belle2}, some new measurements of $\Xi_c(2980)^+$ were presented. In particular, the branching fraction ratio of $B(\Xi_{c}(2980)^{+} \to \Xi_c^{'0} \pi^{+})/B(\Xi_{c}(2815)^{+} \to \Xi_{c}(2645)^{0}\pi^{+},\Xi_{c}(2645)^{0} \to \Xi_c^{+}\pi^{-})\approx 75\% $ was reported, which indicates that $\Xi_c(2980)^+$ decays significantly into $\Xi_c^{'}\pi^+$. The decay of $\Xi_c(2980)^+$ into $\Lambda_c K $ or $ \Xi_c \pi$ mode has not been observed in experiment. In the $^3P_0$ model, decay channels and relevant decay widths of $\Xi_c(2980)^+$ as a $2S$-wave or a $D$-wave excitation are given in Table~\ref{table4} and Table~\ref{table5}, respectively.

\begin{center}
  \begin{table*}
     \caption{Decay widths (MeV) of $ \Xi_c(2980)^+$ as the $2S$-wave excitations.  $\mathcal{B}_2^{'}=B(\Xi_{c}(2980)^{+} \to \Xi_c^{'0} \pi^{+})/B(\Xi_{c}(2815)^{+} \to \Xi_{c}(2645)^{0}\pi^{+},\Xi_{c}(2645)^{0} \to \Xi_c^{+}\pi^{-})$; $\mathcal{B}_2=B(\Xi_{c}(2980)^{+} \to \Xi_c^{'0} \pi^{+})/B(\Xi_{c}(2980)^{+} \to \Xi_{c}(2645)^{0}\pi^{+})$; $\mathcal{B}_3=B(\Xi_{c}(2980)^{+} \to \Sigma_c(2455)^{++} K^-)/B(\Xi_{c}(2980)^{+} \to \Xi_{c}(2645)^{0}\pi^{+})$}
     \begin{tabular}{c |ccccc|cccccccccccccccccc} \hline \hline
 $ \Xi_{cJ_l} (J^P) $ & $ n_{\lambda} $ & $L_\lambda$  & $ n_{\rho} $ & $L_\rho$   & $S_\rho$
 &$\Xi_c\pi    $ &$\Xi_c^{'}\pi $ &$\Xi_c(2645)\pi$   &$\Sigma_c(2455) K$   &$\Lambda_c K$    &$\Gamma_{sum}$    &$\mathcal{B}_2^{'} $ & $\mathcal{B}_2 $ & $\mathcal{B}_3 $ \\
     \hline
  $\acute{\Xi}_{c1}^{'}(\frac{1}{2}^+)$  & 1 &  0 &  0 & 0  &  1  &$0.03\sim3.3$  &$0.4\sim2.8$  &$0.2\sim0.8$    &$0.4\sim0.8$   &$0.09\sim2.7$    &$1.2\sim10$       &$71\sim50\%$   &$1.8\sim3.4$  &$2.0\sim1.0$ \\
  $\acute{\Xi}_{c1}^{'}(\frac{3}{2}^+)$  & 1 &  0 &  0 & 0  &  1  &$0.03\sim3.3$  &$0.1\sim0.7$  &$0.6\sim2.0$    &$0.1\sim0.2$  &$0.09\sim2.7$    &$0.9\sim8.9$      &$21\sim14\%$   &$0.2\sim0.3$  &$0.2\sim0.1$\\
  $\acute{\Xi}_{c0}(\frac{1}{2}^+)$      & 1 &  0 &  0 & 0  &  0  &0              &$0.3\sim2.1$  &$0.7\sim2.4$    &$0.3\sim0.6$   &0                &$1.3\sim5.1$      &$43\sim67\%$   &$0.5\sim0.9$  &$0.5\sim0.3$ \\

  $\tilde{\Xi}_{c1}(\frac{1}{2}^+)$& 0 &0 &  1 & 0  &  1  &$0.3\sim29$    &$3.7\sim25$   &$2.0\sim7.3$    &$4.0\sim7.3$   &$0.8\sim25$      &$11\sim94$      &$71\sim50\%$   &$1.8\sim3.4$  &$2.1\sim1.0$ \\
  $\tilde{\Xi}_{c1}(\frac{3}{2}^+)$& 0 &0 &  1 & 0  &  1  &$0.3\sim29$    &$0.9\sim6.3$  &$5.1\sim18$     &$1.0\sim1.8$   &$0.8\sim25$      &$8.2\sim80$     &$21\sim14\%$   &$0.2\sim0.3$  &$0.2\sim0.1$\\
  $\tilde{\Xi}_{c0}^{'}(\frac{1}{2}^+)$& 0&0&1 & 0  &  0  &0              &$2.8\sim19$   &$6.1\sim22$     &$3.0\sim5.5$   &0                &$12\sim46$    &$43\sim67\%$   &$0.5\sim0.9$    &$0.5\sim0.3$\\

    \hline\hline
    \end{tabular}
    \label{table4}
  \end{table*}
\end{center}

\begin{center}
  \begin{table*}\footnotesize
     \caption{Decay widths (MeV) of $ \Xi_c(2980)^+$ as the $D$-wave excitations.  $\mathcal{B}_2^{'}=B(\Xi_{c}(2980)^{+} \to \Xi_c^{'0} \pi^{+})/B(\Xi_{c}(2815)^{+} \to \Xi_{c}(2645)^{0}\pi^{+},\Xi_{c}(2645)^{0} \to \Xi_c^{+}\pi^{-})$; $\mathcal{B}_2=B(\Xi_{c}(2980)^{+} \to \Xi_c^{'0} \pi^{+})/B(\Xi_{c}(2980)^{+} \to \Xi_{c}(2645)^{0}\pi^{+})$ ; $\mathcal{B}_3=B(\Xi_{c}(2980)^{+} \to  \Sigma_c(2455)^{++} K^-)/B(\Xi_{c}(2980)^{+} \to \Xi_{c}(2645)^{0}\pi^{+})$}
     \begin{tabular}{c |ccccc|cccccccccccccccccc} \hline \hline
 $ \Xi_{cJ_l} (J^P) $ & $ n_{\lambda} $ & $L_\lambda$  & $ n_{\rho} $ & $L_\rho$   & $S_\rho$
 &$\Xi_c\pi    $ &$\Xi_c^{'}\pi $ &$\Xi_c(2645)\pi$   &$\Sigma_c(2455) K$  &$\Lambda_c K$    &$\Gamma_{sum}$   &$\mathcal{B}_2^{'} $ &$\mathcal{B}_2 $ &$\mathcal{B}_3 $\\
     \hline

\label{01}$\Xi_{c1}^{' }(\frac{1}{2}^+)$  & 0 &  2 &  0 & 0  &  1      &$1.8\sim6.2$   &$0.6\sim1.2$    &$0.2\sim0.3$        &$0.3\sim0.3$     &$1.8\sim5.0$     &$4.8\sim13$ &$28\sim18\%$    &$2.7\sim3.6$     &$1.5\sim1.0$\\  
\label{02}$\Xi_{c1}^{' }(\frac{3}{2}^+)$  & 0 &  2 &  0 & 0  &  1      &$1.8\sim6.2$   &$0.2\sim0.3$    &$0.6\sim0.8$        &$0.08\sim0.08$   &$1.8\sim5.0$     &$4.4\sim12$  &$8\sim5\%$      &$0.3\sim0.4$     &$0.2\sim0.1$\\  
\label{03}$\Xi_{c2}^{' }(\frac{3}{2}^+)$  & 0 &  2 &  0 & 0  &  1      &0              &$1.5\sim2.6$    &$0.3\sim0.3$        &$0.7\sim0.7$     &0                &$2.5\sim3.7$  &$113\sim124\%$  &$5.6\sim8.7$     &$3.0\sim2.4$\\  
\label{04}$\Xi_{c2}^{' }(\frac{5}{2}^+)$  & 0 &  2 &  0 & 0  &  1      &0              &$0.1\sim0.03$   &$1.3\sim0.0$        &0                &0                 &$1.4\sim1.8$  &$12\sim2\%$     &$0.08\sim0.02$    &$0$\\  
\label{05}$\Xi_{c3}^{' }(\frac{5}{2}^+)$  & 0 &  2 &  0 & 0  &  1      &$1.0\sim0.3$   &$0.1\sim0.03$   &$0.04\sim0.0$       &0                &$0.6\sim0.2$     &$1.7\sim0.6$  & $14\sim11\%$   &$3.3\sim3.7$      &0 \\  
\label{06}$\Xi_{c3}^{' }(\frac{7}{2}^+)$  & 0 &  2 &  0 & 0  &  1      &$1.0\sim0.3$   &$0.07\sim0.02$  &$0.05\sim0.01$      &0                &$0.6\sim0.2$     &$1.7\sim0.5$  & $8\sim6\%$     &$1.4\sim1.6$      &0\\  
\label{07}$\Xi_{c2}^{  }(\frac{3}{2}^+)$  & 0 &  2 &  0 & 0  &  0      &0              &$1.0\sim1.8$    &$0.2\sim0.2$        &$0.5\sim0.5$     &0                &$1.7\sim2.4$  & $109\sim124\%$ &$4.6\sim8.4$      &$2.5\sim2.3$\\  
\label{08}$\Xi_{c2}^{  }(\frac{5}{2}^+)$  & 0 &  2 &  0 & 0  &  0      &0              &$0.2\sim0.04$   &$0.9\sim1.2$        &0                &0                &$1.1\sim1.2$  & $24\sim5\%$    &$0.2\sim0.04$     &$0$\\  
\label{09}$\hat\Xi_{c1}^{' }(\frac{1}{2}^+)$ & 0 &  0 &  0 & 2  &  1   &$16\sim55$     &$5.8\sim11$     &$2.2\sim2.9$        &$3.0\sim2.8$     &$16\sim44$       &$42\sim115$ & $28\sim18\%$  &$2.7\sim3.6$         &$1.5\sim1.0$\\  
\label{10}$\hat\Xi_{c1}^{' }(\frac{3}{2}^+)$ & 0 &  0 &  0 & 2  &  1   &$16\sim55$     &$1.5\sim2.6$    &$5.4\sim7.3$        &$0.7\sim0.7$     &$16\sim44$       &$39\sim110$ &$8\sim5\%$     &$0.3\sim0.4$         &$0.2\sim0.1$\\  
\label{11}$\hat\Xi_{c2}^{' }(\frac{3}{2}^+)$ & 0 &  0 &  0 & 2  &  1   &0              &$13.1\sim24$    &$2.4\sim2.7$        &$6.7\sim6.2$     &0                &$22\sim33$ &$112\sim124\%$ &$5.6\sim8.7$          &$3.0\sim2.4$  \\  
\label{12}$\hat\Xi_{c2}^{' }(\frac{5}{2}^+)$ & 0 &  0 &  0 & 2  &  1   &0              &$1.0\sim0.3$    &$12\sim16$          &0                &0                &$13\sim16$ &$12\sim3\%$    &$0.08\sim0.02$        &0\\  
\label{13}$\hat\Xi_{c3}^{' }(\frac{5}{2}^+)$ & 0 &  0 &  0 & 2  &  1   &$8.7\sim3.1$   &$1.1\sim0.3$    &$0.3\sim0.08$       &0                &$5.5\sim1.8$     &$16\sim5.2$  &$14\sim11\%$   &$3.3\sim3.7$        &0\\  
\label{14}$\hat\Xi_{c3}^{' }(\frac{7}{2}^+)$ & 0 &  0 &  0 & 2  &  1   &$8.7\sim3.1$   &$0.6\sim0.2$    &$0.5\sim0.1$        &0                &$5.5\sim1.8$     &$15\sim5.1$  &$8\sim6\%$     &$1.4\sim1.6$        &0\\  
\label{15}$\hat\Xi_{c2}^{ }(\frac{3}{2}^+)$  & 0 &  0 &  0 & 2  &  0   &0              &$8.7\sim16$     &$1.9\sim1.9$        &$4.5\sim4.2$     &0                &$15\sim22$ &$109\sim124\%$ &$4.5\sim8.3$          &$2.4\sim2.3$ \\  
\label{16}$\hat\Xi_{c2}^{ }(\frac{5}{2}^+)$  & 0 &  0 &  0 & 2  &  0   &0              &$1.5\sim0.4$    &$8.1\sim11$         &0                &0                &$9.5\sim11$  &$24\sim6\%$    &$0.2\sim0.04$       &0\\  
\label{17}$\check\Xi_{c0}^{'0}(\frac{1}{2}^+)$ & 0 &  1 &  0 & 1  &  0 &0              &$1.9\sim13$     &$4.1\sim15$         &$2.0\sim3.7$     &0                &$8.0\sim31$  &$43\sim67\%$   &$0.5\sim0.9$        &$0.5\sim0.3$\\  
\label{18}$\check\Xi_{c1}^{'1}(\frac{1}{2}^+)$ & 0 &  1 &  0 & 1  &  0 &0              &0               &0                   &0                &0       &0    &$\cdots$ &$\cdots$ &$\cdots$\\  
\label{19}$\check\Xi_{c1}^{'1}(\frac{3}{2}^+)$ & 0 &  1 &  0 & 1  &  0 &0              &0               &0                   &0                &0       &0    &$\cdots$ &$\cdots$ &$\cdots$\\  
\label{20}$\check\Xi_{c2}^{'2}(\frac{3}{2}^+)$ & 0 &  1 &  0 & 1  &  0 &0              &$5.8\sim11$     &$1.3\sim1.3$        &$3.0\sim2.8$     &0                &$10\sim15$  &$109\sim124\%$  &$4.5\sim8.3$    &$2.4\sim2.3$\\  
\label{21}$\check\Xi_{c2}^{'2}(\frac{5}{2}^+)$ & 0 &  1 &  0 & 1  &  0 &0              &$1.0\sim0.3$    &$5.4\sim7.1$        &0                &0                &$6.4\sim7.3$   &$24\sim6\%$     &$0.2\sim0.04$    &0\\  
\label{22}$\check\Xi_{c1}^{\ 0}(\frac{1}{2}^+)$& 0 &  1 &  0 & 1  &  1 &$0.2\sim20$    &$2.5\sim17$     &$1.4\sim4.9$        &$2.7\sim4.9$     &$0.6\sim16$     &$7.3\sim62$  &$71\sim50\%$    &$1.8\sim3.4$    &$2.1\sim1.1$\\  
\label{23}$\check\Xi_{c1}^{\ 0}(\frac{3}{2}^+)$& 0 &  1 &  0 & 1  &  1 &$0.2\sim20$    &$0.6\sim4.2$    &$3.4\sim12$         &$0.7\sim1.2$     &$0.6\sim16$     &$5.5\sim54$  &$21\sim14\%$    &$0.2\sim0.3$    &$0.2\sim0.1$\\  
\label{24}$\check\Xi_{c0}^{\ 1}(\frac{1}{2}^+)$ & 0 &  1 &  0 & 1  &  1&0              &0             &0        &0     &0     &0    &$\cdots$ &$\cdots$ &$\cdots$\\  
\label{25}$\check\Xi_{c1}^{\ 1}(\frac{1}{2}^+)$ & 0 &  1 &  0 & 1  &  1&0              &0             &0        &0     &0     &0    &$\cdots$ &$\cdots$ &$\cdots$\\  
\label{26}$\check\Xi_{c1}^{\ 1}(\frac{3}{2}^+)$ & 0 &  1 &  0 & 1  &  1&0              &0             &0        &0     &0     &0    &$\cdots$ &$\cdots$ &$\cdots$\\  
\label{27}$\check\Xi_{c2}^{\ 1}(\frac{3}{2}^+)$ & 0 &  1 &  0 & 1  &  1&0              &0             &0        &0     &0     &0    &$\cdots$ &$\cdots$ &$\cdots$\\  
\label{28}$\check\Xi_{c2}^{\ 1}(\frac{5}{2}^+)$ & 0 &  1 &  0 & 1  &  1&0              &0             &0        &0     &0     &0    &$\cdots$ &$\cdots$ &$\cdots$\\  
\label{29}$\check\Xi_{c1}^{\ 2}(\frac{1}{2}^+)$ & 0 &  1 &  0 & 1  &  1&$11\sim37$     &$3.9\sim7.0$    &$1.4\sim2.0$         &$2.0\sim1.9$     &$11\sim30$     &$28\sim77$  &$28\sim18\%$   &$2.7\sim3.6$    &$1.5\sim1.0$\\  
\label{30}$\check\Xi_{c1}^{\ 2}(\frac{3}{2}^+)$ & 0 &  1 &  0 & 1  &  1&$11\sim37$     &$1.0\sim1.8$    &$3.6\sim4.9$         &$0.5\sim0.5$     &$11\sim30$     &$26\sim73$  &$8\sim5\%$     &$0.3\sim0.4$    &$0.15\sim0.1$\\  
\label{31}$\check\Xi_{c2}^{\ 2}(\frac{3}{2}^+)$ & 0 &  1 &  0 & 1  &  1&0              &$8.7\sim16$     &$1.6\sim1.8$         &$4.5\sim4.2$     &0              &$15\sim22$  &$112\sim124\%$ &$5.6\sim8.7$    &$3.0\sim2.4$ \\  
\label{32}$\check\Xi_{c2}^{\ 2}(\frac{5}{2}^+)$ & 0 &  1 &  0 & 1  &  1&0              &$0.6\sim0.2$    &$7.9\sim11$          &0                &0              &$8.6\sim11$   &$12\sim3\%$    &$0.08\sim0.02$    &0\\  
\label{33}$\check\Xi_{c3}^{\ 2}(\frac{5}{2}^+)$ & 0 &  1 &  0 & 1  &  1&$5.8\sim2.0$   &$0.7\sim0.2$    &$0.2\sim0.05$        &0                &$3.7\sim1.2$   &$10\sim3.5$    &$14\sim11\%$   &$3.3\sim3.7$    &0\\  
\label{34}$\check\Xi_{c3}^{\ 2}(\frac{7}{2}^+)$ & 0 &  1 &  0 & 1  &  1&$5.8\sim2.0$   &$0.4\sim0.1$    &$0.3\sim0.07$        &0                &$3.7\sim1.2$   &$10\sim3.4$    &$8\sim6\%$     &$1.4\sim1.6$    &0\\  
    \hline\hline
    \end{tabular}
    \label{table5}
  \end{table*}
\end{center}

From Table~\ref{table4}, assignment of $\Xi_c(2980)^+$ with $\acute{\Xi}_{c1}^{'}(\frac{3}{2}^+)$ or $\tilde{\Xi}_{c1}(\frac{3}{2}^+)$ seems not reasonable for a large deviation of the branching fraction ratio $\mathcal{B}_2^{'}$ between theory and experiment. Assignment of $\Xi_c(2980)^+$ with $\acute{\Xi}_{c1}^{'}(\frac{1}{2}^+)$ or $\acute{\Xi}_{c0}(\frac{1}{2}^+)$ could be excluded either for a narrow decay widths. $\Xi_c(2980)^+$ is very possibly a radial excitation $\tilde{\Xi}_{c1}(\frac{1}{2}^+)$ or $\tilde{\Xi}_{c0}^{'}(\frac{1}{2}^+)$ with $J^P={1\over 2}^+$. In the case as a $\tilde{\Xi}_{c0}^{'}(\frac{1}{2}^+)$ with $J^P={1\over 2}^+$, the decay mode $\Lambda_cK$ vanishes. In these assignment, $ \Xi_c(2980)^+$ has a total decay width and branching fraction ratio $B'_2$ consistent with experimental data from Belle collaboration.

From the branching fraction ratio $B'_2$ in Table~\ref{table5}, $\Xi_c(2980)^+$ can only be assigned as a $\check\Xi_{c0}^{'0}(\frac{1}{2}^+)$ or $\check\Xi_{c1}^{\ 0}(\frac{1}{2}^+)$ with $J^P={1\over 2}^+$ in the case as a $D$-wave excitation. In these assignments, $ \Xi_c(2980)^+$ has also a total decay width comparable with experiment.

In Ref.~\cite{ye1}, it was indicated that $\Xi_c(2980)^+$ could be identified as a $P$-wave excitation $\Xi_{c1}^{'}(\frac{1}{2}^-)$ with $J^P={1\over 2}^-$. Obviously, there are three possible assignments for $\Xi_c(2980)^+$ once the $2S$-wave and $D$-wave possibilities have been taken into account. However, the ratios $\mathcal{B}_2=B(\Xi_{c}(2980)^{+} \to \Xi_c^{'0} \pi^{+})/B(\Xi_{c}(2980)^{+} \to \Xi_{c}(2645)^{0}\pi^{+})$ and $\mathcal{B}_3=B(\Xi_{c}(2980)^{+} \to  \Sigma_c(2455)^{++} K^-)/B(\Xi_{c}(2980)^{+} \to \Xi_{c}(2645)^{0}\pi^{+})$ are different in these three different assignments. As shown in Ref.~\cite{ye1}, the ratios $\mathcal{B}_2$ and $\mathcal{B}_3$ of $\Xi_c(2980)$ as $\Xi_{c1}^{'}(\frac{1}{2}^-)$ are $7.0\sim77$ and $36\sim144$, respectively. From Tables~\ref{table4}-\ref{table5}, the ratios $\mathcal{B}_2$ and $\mathcal{B}_3$ of $\Xi_c(2980)$ as $\check{\Xi}_{c0}^{'}(\frac{1}{2}^+)$ are $0.5\sim0.9$ and $0.5\sim 0.3$, respectively. The ratios $\mathcal{B}_2$ and $\mathcal{B}_3$ of $\Xi_c(2980)$ as $\check{\Xi}_{c1}(\frac{1}{2}^+)$ are $1.8\sim3.4$ and $2.1\sim1.1$, respectively. Obviously, the measurements of these ratios such as $\mathcal{B}_2$ or $\mathcal{B}_3$ are important for the identification of $\Xi_c(2980)^+$ in the future.

\subsubsection{$\Xi_c(3055)$}

$\Xi_c(3055)^+$ with a total decay width $\Gamma_{\Xi_{c}(3055)^{+}}=(7.8 \pm 1.9)$ MeV was observed in the $\Sigma_c(2455)^{++}K^- $ and $\Lambda D^+$ channels~\cite{belle3}. In particular, the ratio of branching fraction was also measured in experiment, $$\frac{\Gamma_{\Xi_{c}(3055)^{+} \to \Lambda D^{+}}}{\Gamma_{\Xi_{c}(3055)^{+} \to \Sigma_c(2455)^{++} K^-)}}=5.09\pm1.01(stat)\pm0.76(sys).$$

If $\Xi_c(3055)^+$ is considered as a $2S$-wave excitation, the numerical results are presented in Table~\ref{table6}.
\begin{center}
  \begin{table*}\tiny
     \caption{Decay widths (MeV) of $ \Xi_c(3055)^+$ as the  $2S$-wave excitation in different assignments. $\mathcal{B}_4=B(\Xi_{c}(3055)^{+} \to \Sigma_c(2520)^{++} K^-)/B(\Xi_{c}(3055)^{+} \to \Sigma_{c}(2455)^{++} K^-)$;
     $\mathcal{B}_5=B(\Xi_{c}(3055)^{+} \to \Xi_c(2645)^{0} \pi^+)/B(\Xi_{c}(3055)^{+} \to \Sigma_{c}(2455)^{++} K^-)$;
     $\mathcal{B}_6=B(\Xi_{c}(3055)^{+} \to \Lambda D^+)/B(\Xi_{c}(3055)^{+} \to \Sigma_{c}(2455)^{++} K^-)$}
     \begin{tabular}{c |ccccc|cccccccccccccccccc} \hline \hline
 $ \Xi_{cJ_l} (J^P) $ & $ n_{\lambda} $ & $L_\lambda$  & $ n_{\rho} $ & $L_\rho$   & $S_\rho$
 &$\Xi_c\pi        $ &$\Xi_c^{'}\pi $  &$\Xi_c(2645)\pi$       &$\Sigma_c(2455) K$        &$\Sigma_c(2520) K $
 &$\Lambda_c K$      &$\Lambda D^+$    &$\Gamma_{sum}$     & $\mathcal{B}_4 $ &  $\mathcal{B}_5 $  &$\mathcal{B}_6  $  \\
     \hline

\label{01}$\acute{\Xi}_{c1}^{'}(\frac{1}{2}^+)$  & 1 &  0 &  0 & 0   &  1      &$0.1\sim3.5$   &$0.1\sim4.1$    &$0.2\sim1.6$        &$1.5\sim6.8$     &$0.3\sim0.7$     &$0.09\sim2.9$    &$0.6\sim3.5$    &$2.8\sim23$    &$0.2\sim0.1$    &$0.1\sim0.2$   &$0.6\sim0.8$\\  

\label{02} $\acute{\Xi}_{c1}^{'}(\frac{3}{2}^+)$  & 1 &  0 &  0 & 0  &  1      &$0.1\sim3.5$   &$0.03\sim1.0$   &$0.5\sim3.9$        &$0.4\sim1.7$     &$0.7\sim1.7$     &$0.09\sim2.9$    &$2.2\sim14$     &$2.8\sim29$    &$2.0\sim1.0$    &$1.2\sim2.3$   &$9.0\sim12$\\  

\label{03}$\acute{\Xi}_{c0}(\frac{1}{2}^+)$      & 1 &  0 &  0 & 0   &  0      &0              &$0.08\sim3.1$   &$0.5\sim4.7$        &$1.1\sim5.1$     &$0.8\sim2.0$      &0                &$1.6\sim11$    &$4.2\sim25$    &$0.8\sim0.4$    &$0.5\sim0.9$   &$2.3\sim3.1$    \\  

\label{04}$\tilde{\Xi}_{c1}(\frac{1}{2}^+)$& 0 &0 &  1 & 0   &  1      &$1.1\sim32$    &$1.0\sim37$     &$1.6\sim14$        &$13\sim61$       &$2.5\sim6.0$      &$0.8\sim26$      &0              &$20\sim176$    &$0.2\sim0.1$     &$0.1\sim0.2$     &$0$\\  

\label{05}$\tilde{\Xi}_{c1}(\frac{3}{2}^+)$& 0 &0 &  1 & 0   &  1      &$1.1\sim32$    &$0.3\sim9.3$    &$4.1\sim35$       &$3.3\sim15$      &$6.3\sim15$       &$0.8\sim26$      &0              &$16\sim133$    &$2.0\sim1.0$    &$1.2\sim2.3$   &0 \\  

\label{06}$\tilde{\Xi}_{c0}^{'}(\frac{1}{2}^+)$& 0&0&1 & 0   &  0      &0              &$0.8\sim28$    &$4.9\sim42$      &$9.9\sim46$      &$7.6\sim18$       &0                &0              &$23\sim134$    &$0.8\sim0.4$   &$0.5\sim0.9$   &0 \\  

    \hline\hline
    \end{tabular}
    \label{summary_simfit}
    \label{table6}
  \end{table*}
\end{center}

Once the theoretical results from Table~\ref{table6} are compared with experimental data under some uncertainties, $\Xi_c(3055)^+$ is possibly a $\acute{\Xi}_{c1}^{'}(\frac{3}{2}^+)$ or $\acute{\Xi}_{c0}(\frac{1}{2}^+)$. However, the features of their decays into $\Xi_c \pi$ and $\Lambda_c K$ modes are different in these two different assignments. In the case of $\acute{\Xi}_{c1}^{'}(\frac{3}{2}^+)$, the modes $\Xi_c \pi$ and $\Lambda_c K$ may be observed while these modes vanish in the case of $\acute{\Xi}_{c0}(\frac{1}{2}^+)$.

If $\Xi_c(3055)^+$ is considered as a $D$-wave excitation, the results are given in Table~\ref{table7}.
\begin{center}
  \begin{table*}\tiny
     \caption{Decay widths (MeV) of $ \Xi_c(3055)^+$ as the $D$-wave excitation in different assignments. $\mathcal{B}_4=B(\Xi_{c}(3055)^{+} \to \Sigma_c(2520)^{++} K^-)/B(\Xi_{c}(3055)^{+} \to \Sigma_{c}(2455)^{++} K^-)$;
     $\mathcal{B}_5=B(\Xi_{c}(3055)^{+} \to \Xi_c(2645)^{0} \pi^+)/B(\Xi_{c}(3055)^{+} \to \Sigma_{c}(2455)^{++} K^-)$;
     $\mathcal{B}_6=B(\Xi_{c}(3055)^{+} \to \Lambda D^+)/B(\Xi_{c}(3055)^{+} \to \Sigma_{c}(2455)^{++} K^-)$}
     \begin{tabular}{c |ccccc|cccccccccccccccccc} \hline \hline
 $ \Xi_{cJ_l} (J^P) $ & $ n_{\lambda} $ & $L_\lambda$  & $ n_{\rho} $ & $L_\rho$   & $S_\rho$
 &$\Xi_c\pi        $ &$\Xi_c^{'}\pi $  &$\Xi_c(2645)\pi$       &$\Sigma_c(2455) K$        &$\Sigma_c(2520) K $
 &$\Lambda_c K$      &$\Lambda D^+$    &$\Gamma_{sum}$     & $\mathcal{B}_4 $ &  $\mathcal{B}_5 $  &$\mathcal{B}_6  $   \\
     \hline

\label{01}$\Xi_{c1}^{' }(\frac{1}{2}^+)$  & 0 &  2 &  0 & 0  &  1      &$1.0\sim7.6$   &$0.6\sim1.9$    &$0.3\sim0.7$        &$1.8\sim2.8$     &$0.4\sim0.4$     &$0.8\sim6.3$     &$19\sim43$      &$24\sim62$     &$0.2\sim0.1$    &$0.2\sim0.2$   &$15\sim23$    \\  

\label{02}$\Xi_{c1}^{' }(\frac{3}{2}^+)$  & 0 &  2 &  0 & 0  &  1      &$1.0\sim7.6$   &$0.2\sim0.5$    &$0.8\sim1.7$        &$0.5\sim0.7$     &$1.0\sim1.0$     &$0.8\sim6.3$     &$1.2\sim2.7$    &$5.4\sim20$    &$2.2\sim1.5$    &$1.9\sim2.4$   &$3.9\sim5.7$    \\  

\label{03}$\Xi_{c2}^{' }(\frac{3}{2}^+)$  & 0 &  2 &  0 & 0  &  1      &0              &$1.4\sim4.3$    &$0.6\sim0.7$        &$4.1\sim6.3$     &$0.4\sim0.4$      &0                &$10\sim24$     &$17\sim36$     &$0.1\sim0.1$    &$0.1\sim0.1$   &$3.9\sim5.7$    \\  

\label{04}$\Xi_{c2}^{' }(\frac{5}{2}^+)$  & 0 &  2 &  0 & 0  &  1      &0              &$0.4\sim0.1$    &$1.9\sim3.7$        &$0.2\sim0.0$     &$2.1\sim2.2$      &0                &$0.4\sim0.1$   &$5.6\sim6.1$   &$13\sim53$      &$11\sim86$     &$3.7\sim4.6$    \\  

\label{05}$\Xi_{c3}^{' }(\frac{5}{2}^+)$  & 0 &  2 &  0 & 0  &  1      &$2.2\sim0.9$   &$0.4\sim0.1$    &$0.2\sim0.06$       &$0.2\sim0.05$    &$0.01\sim0.0$     &$1.7\sim0.7$     &$0.03\sim0.01$ &$4.8\sim1.9$   &$0.1\sim0.0$    &$1.1\sim1.2$    &$0.2\sim0.3$    \\  

\label{06}$\Xi_{c3}^{' }(\frac{7}{2}^+)$  & 0 &  2 &  0 & 0  &  1      &$2.2\sim0.9$   &$0.3\sim0.08$  &$0.3\sim0.08$      &$0.1\sim0.03$    &$0.01\sim0.0$     &$1.7\sim0.7$     &$1.1\sim0.3$   &$5.6\sim2.2$   &$0.1\sim0.1$   &$2.5\sim2.8$     &$15\sim19$    \\  

\label{07}$\Xi_{c2}^{  }(\frac{3}{2}^+)$  & 0 &  2 &  0 & 0  &  0      &0              &$1.0\sim2.9$    &$0.6\sim0.5$        &$2.7\sim4.2$     &$0.3\sim0.3$      &0                &$7.0\sim16$    &$12\sim24$     &$0.1\sim0.1$    &$0.2\sim0.1$   &$3.9\sim5.7$ \\  

\label{08}$\Xi_{c2}^{  }(\frac{5}{2}^+)$  & 0 &  2 &  0 & 0  &  0      &0              &$0.6\sim0.2$    &$1.4\sim2.5$        &$0.3\sim0.06$    &$1.4\sim1.5$      &0                &$0.6\sim0.2$   &$4.3\sim4.4$   &$5.6\sim24$     &$5.4\sim39$    &$3.7\sim4.6$ \\  

\label{09}$\hat\Xi_{c1}^{' }(\frac{1}{2}^+)$ & 0 &  0 &  0 & 2  &  1   &$8.4\sim67$    &$5.7\sim17$     &$3.0\sim6.0$        &$16\sim25$       &$3.5\sim3.6$      &$7.4\sim56$      &0              &$44\sim174$    &$0.2\sim0.1$    &$0.2\sim0.2$  &0 \\  

\label{10}$\hat\Xi_{c1}^{' }(\frac{3}{2}^+)$ & 0 &  0 &  0 & 2  &  1   &$8.3\sim67$    &$1.4\sim4.2$    &$7.5\sim15$        &$4.1\sim6.2$     &$8.7\sim9.1$      &$7.4\sim56$      &0              &$38\sim157$    &$2.2\sim1.5$    &$1.9\sim2.4$  &0\\  

\label{11}$\hat\Xi_{c2}^{' }(\frac{3}{2}^+)$ & 0 &  0 &  0 & 2  &  1   &0              &$13\sim38$      &$5.1\sim6.1$        &$37\sim56$       &$3.3\sim3.3$       &0                &0              &$58\sim103$    &$0.1\sim0.1$    &$0.1\sim0.1$  &0\\  

\label{12}$\hat\Xi_{c2}^{' }(\frac{5}{2}^+)$ & 0 &  0 &  0 & 2  &  1   &0              &$3.5\sim1.2$    &$17\sim33$          &$1.5\sim0.4$     &$19\sim20$         &0               &0              &$41\sim54$     &$12\sim51$      &$11\sim81$  &0\\  

\label{13}$\hat\Xi_{c3}^{' }(\frac{5}{2}^+)$ & 0 &  0 &  0 & 2  &  1   &$20\sim8.8$    &$4.0\sim1.3$    &$1.9\sim0.6$       &$1.8\sim0.4$     &$0.1\sim0.02$     &$16\sim6.9$      &0              &$43\sim18$     &$0.1\sim0.0$    &$1.1\sim1.2$ &0\\  

\label{14}$\hat\Xi_{c3}^{' }(\frac{7}{2}^+)$ & 0 &  0 &  0 & 2  &  1   &$20\sim8.8$    &$2.2\sim0.7$    &$2.6\sim0.7$        &$1.0\sim0.3$     &$0.1\sim0.03$     &$16\sim6.9$      &0              &$41\sim17$     &$0.1\sim0.1$    &$2.5\sim2.8$ &0\\  

\label{15}$\hat\Xi_{c2}^{ }(\frac{3}{2}^+)$  & 0 &  0 &  0 & 2  &  0   &0              &$8.6\sim25$     &$5.4\sim4.6$        &$24\sim37$       &$2.3\sim2.2$      &0                &0              &$41\sim70$     &$0.1\sim0.1$    &$0.2\sim0.1$  &0\\  

\label{16}$\hat\Xi_{c2}^{ }(\frac{5}{2}^+)$  & 0 &  0 &  0 & 2  &  0   &0              &$5.2\sim1.7$    &$12\sim22$         &$2.3\sim0.6$    &$13\sim13$         &0                &0              &$33\sim37$     &$5.5\sim23$     &$5.3\sim37$   &0\\  

\label{17}$\check\Xi_{c0}^{'0}(\frac{1}{2}^+)$ & 0 &  1 &  0 & 1  &  0 &0              &$0.5\sim19$     &$3.3\sim28$         &$6.6\sim31$     &$8.7\sim19$        &0                &0              &$19\sim96$     &$1.3\sim0.6$    &$0.5\sim0.9$  &0\\  

\label{18}$\check\Xi_{c1}^{'1}(\frac{1}{2}^+)$ & 0 &  1 &  0 & 1  &  0 &0              &0             &0                 &0              &0       &0      &0       &0    &$\cdots$ &$\cdots$ &$\cdots$\\  
\label{19}$\check\Xi_{c1}^{'1}(\frac{3}{2}^+)$ & 0 &  1 &  0 & 1  &  0 &0              &0             &0                 &0              &0       &0      &0       &0    &$\cdots$ &$\cdots$ &$\cdots$\\  

\label{20}$\check\Xi_{c2}^{'2}(\frac{3}{2}^+)$ & 0 &  1 &  0 & 1  &  0 &0              &$5.8\sim17$     &$3.6\sim3.1$        &$16\sim25$       &$1.5\sim1.5$     &0                &0               &$27\sim46$     &$0.1\sim0.1$    &$0.2\sim0.1$  &0\\  

\label{21}$\check\Xi_{c2}^{'2}(\frac{5}{2}^+)$ & 0 &  1 &  0 & 1  &  0 &0              &$3.5\sim1.2$    &$8.3\sim15$        &$1.5\sim0.4$     &$8.4\sim8.7$     &0                &0               &$22\sim25$     &$5.5\sim23$     &$5.3\sim37$   &0\\  

\label{22}$\check\Xi_{c1}^{\ 0}(\frac{1}{2}^+)$& 0 &  1 &  0 & 1  &  1 &$0.8\sim21$    &$0.7\sim25$     &$1.1\sim9.4$        &$8.8\sim41$      &$2.9\sim6.3$     &$0.5\sim18$      &0               &$15\sim120$    &$0.3\sim0.2$    &$0.1\sim0.2$   &0\\  

\label{23}$\check\Xi_{c1}^{\ 0}(\frac{3}{2}^+)$& 0 &  1 &  0 & 1  &  1 &$0.8\sim21$    &$0.2\sim6.2$    &$2.7\sim23$         &$2.2\sim10$      &$7.2\sim16$     &$0.5\sim17$      &0                &$14\sim94$     &$3.4\sim1.6$    &$1.2\sim2.3$  &0\\  

\label{24}$\check\Xi_{c0}^{\ 1}(\frac{1}{2}^+)$ & 0 &  1 &  0 & 1  &  1&0              &0             &0        &0     &0     &0    &0       &0    &$\cdots$ &$\cdots$ &$\cdots$\\  

\label{25}$\check\Xi_{c1}^{\ 1}(\frac{1}{2}^+)$ & 0 &  1 &  0 & 1  &  1&0              &0             &0        &0     &0     &0    &0       &0    &$\cdots$ &$\cdots$ &$\cdots$\\  
\label{26}$\check\Xi_{c1}^{\ 1}(\frac{3}{2}^+)$ & 0 &  1 &  0 & 1  &  1&0              &0             &0        &0     &0     &0    &0       &0    &$\cdots$ &$\cdots$ &$\cdots$\\  
\label{27}$\check\Xi_{c2}^{\ 1}(\frac{3}{2}^+)$ & 0 &  1 &  0 & 1  &  1&0              &0             &0        &0     &0     &0    &0       &0    &$\cdots$ &$\cdots$ &$\cdots$\\  
\label{28}$\check\Xi_{c2}^{\ 1}(\frac{5}{2}^+)$ & 0 &  1 &  0 & 1  &  1&0              &0             &0        &0     &0     &0    &0       &0    &$\cdots$ &$\cdots$ &$\cdots$\\  

\label{29}$\check\Xi_{c1}^{\ 2}(\frac{1}{2}^+)$ & 0 &  1 &  0 & 1  &  1&$5.6\sim45$    &$3.8\sim11$     &$2.0\sim4.0$         &$11\sim17$       &$2.3\sim2.4$   &$4.9\sim37$     &0                  &$30\sim116$    &$0.2\sim0.1$    &$0.2\sim0.2$  &0\\  

\label{30}$\check\Xi_{c1}^{\ 2}(\frac{3}{2}^+)$ & 0 &  1 &  0 & 1  &  1&$5.6\sim45$    &$1.0\sim2.8$    &$5.0\sim10$         &$2.7\sim4.2$     &$5.8\sim6.1$   &$4.9\sim37$     &0                  &$25\sim105$    &$2.2\sim1.5$    &$1.9\sim2.4$  &0\\  

\label{31}$\check\Xi_{c2}^{\ 2}(\frac{3}{2}^+)$ & 0 &  1 &  0 & 1  &  1&0              &$8.6\sim25$     &$3.4\sim4.1$         &$24\sim37$       &$2.2\sim2.2$   &0               &0                  &$39\sim69$     &$0.1\sim0.1$    &$0.1\sim0.1$  &0\\  

\label{32}$\check\Xi_{c2}^{\ 2}(\frac{5}{2}^+)$ & 0 &  1 &  0 & 1  &  1&0              &$2.3\sim0.8$    &$12\sim22$          &$1.0\sim0.3$     &$13\sim13$      &0               &0                  &$28\sim36$    &$12\sim51$      &$11\sim81$  &0\\  

\label{33}$\check\Xi_{c3}^{\ 2}(\frac{5}{2}^+)$ & 0 &  1 &  0 & 1  &  1&$13\sim5.8$    &$2.7\sim0.9$    &$1.3\sim0.4$        &$1.2\sim0.3$     &$0.1\sim0.01$   &$11\sim4.6$    &0                  &$29\sim12$     &$0.1\sim0.0$    &$1.1\sim1.2$ &0\\  

\label{34}$\check\Xi_{c3}^{\ 2}(\frac{7}{2}^+)$ & 0 &  1 &  0 & 1  &  1&$13\sim5.8$    &$1.5\sim0.5$    &$1.7\sim0.5$        &$0.7\sim0.2$     &$0.1\sim0.02$   &$11\sim4.6$    &0                  &$28\sim12$     &$0.1\sim0.1$    &$2.5\sim2.8$    &0\\  
    \hline\hline
    \end{tabular}
    \label{summary_simfit}
    \label{table7}
  \end{table*}
\end{center}

From this table, the $26$ types of configurations with vanish $\Lambda D^+$ channel seem impossible for $\Xi_c(3055)^+$. $\Xi_{c1}^{' }(\frac{1}{2}^+)$ or $\Xi_{c2}^{' }(\frac{3}{2}^+)$ seems impossible for the large predicted total width. $\Xi_{c3}^{' }(\frac{5}{2}^+)$ or $\Xi_{c3}^{' }(\frac{7}{2}^+)$ seems impossible either for a too much small or large ratio $\mathcal{B}_6=B(\Xi_{c}(3055)^{+} \to \Lambda D^+)/B(\Xi_{c}(3055)^{+} \to \Sigma_{c}(2455)^{++} K^-)$ in comparison with experiment. $\Xi_c(3055)^+$ may be a $\Xi_{c1}^{' }(\frac{3}{2}^+)$, $\Xi_{c2}^{' }(\frac{5}{2}^+)$, $\Xi_{c2}^{  }(\frac{3}{2}^+)$ or $\Xi_{c2}^{  }(\frac{5}{2}^+)$. In the case of $\Xi_{c1}^{' }(\frac{3}{2}^+)$, $\Xi_c(3055)^+$ has a non-vanish $\Xi_c \pi$ or $\Lambda_c K$ mode, while these modes vanish in other three assignments. In particular, $\Xi_c(3055)^+$ as a $\Xi_{c2}^{  }(\frac{3}{2}^+)$ decays mainly into $\Sigma_c(2455)^{++} K^- $ and $\Lambda D^+$ modes, which is also consistent with the experiments. Similar conclusions were obtained in a chiral quark model~\cite{zhong} and from the EHQ decay formula~\cite{chen}.

The ratios $\mathcal{B}_4=B(\Xi_{c}(3055)^{+} \to \Sigma_c(2520)^{++} K^-)/B(\Xi_{c}(3055)^{+} \to \Sigma_{c}(2455)^{++} K^-)$ and $\mathcal{B}_5=B(\Xi_{c}(3055)^{+} \to \Xi_c(2645)^{0} \pi^+)/B(\Xi_{c}(3055)^{+} \to \Sigma_{c}(2455)^{++} K^-)$ of $\Xi_c(3055)^+$ are obviously different in these four possible assignments, which could be employed to identify $\Xi_c(3055)^+$ in forthcoming experiments.

In Ref.~\cite{zhang4}, $\Xi_c(3055)^+$ was assigned with a $\hat\Xi_{c3}^\prime(\frac{5}{2}^+)$ or $\check\Xi_{c3}^2(\frac{5}{2}^+)$ with $J^P={5\over 2}^+$, it was also assigned with a $\hat\Xi_{c3}^\prime(\frac{7}{2}^+)$ or $\check\Xi_{c3}^2(\frac{7}{2}^+)$ with $J^P={7\over 2}^+$. As indicated in Eqs.~[\ref{eq1}-\ref{eq3}], there are three possible rearrangements for the decay process of a $\Xi_c$ baryon. In the processes $(1-2)$, the excited charmed baryon decays into a charmed baryon and a light meson, while the excited charmed baryon decays into a light baryon and a charmed meson in process $(3)$. In Ref.~\cite{zhang4}, these three processes are not dealt with in a proper way. When these three processes are dealt with differently, the decay width of the $\Lambda D^+$ mode vanishes. Therefore, the assignment of $\Xi_c(3055)^+$ with a $\hat\Xi_{c3}^\prime(\frac{5}{2}^+)$ or $\check\Xi_{c3}^2(\frac{5}{2}^+)$ with $J^P={5\over 2}^+$ is not suitable, and the assignment of $\Xi_c(3055)^+$ with a $\hat\Xi_{c3}^\prime(\frac{7}{2}^+)$ or $\check\Xi_{c3}^2(\frac{7}{2}^+)$ with $J^P={7\over 2}^+$ is not suitable either.

\subsubsection{$\Xi_c(3080)$}

$\Xi_c(3080)^+$ with a width $\Gamma_{\Xi_{c}(3080)^{+}}=(3.6 \pm 1.1)$ MeV was observed in the $\Sigma_c(2455)^{++} K^- $, $\Sigma_c(2520)^{++} K^- $ and $\Lambda D^+$ decay channels~\cite{Chistov:0606051,belle4}, but was not observed in the $\Lambda_c K$ channel.
In addition, the following ratios of branching fractions were obtained by Belle~\cite{belle4} collaboration,
$$\frac{\Gamma_{\Xi_{c}(3080)^{+} \to \Lambda D^{+}}}{\Gamma_{\Xi_{c}(3080)^{+} \to \Sigma_c(2455)^{++} K^-)}}=1.29\pm0.30(stat)\pm0.15(sys),$$
$$\frac{\Gamma_{\Xi_{c}(3080)^{+} \to \Sigma_{c}(2520)^{++}}K^{-} }{\Gamma_{\Xi_{c}(3080)^{+} \to \Sigma_c(2455)^{++} K^-)}}=1.07\pm0.27(stat)\pm0.04(sys).$$

Numerical results of $\Xi_c(3080)$ as a $2S$-wave excitation are given in Table~\ref{table8}. From Table~\ref{table8}, the three possibilities with vanish predicted $\mathcal{B}_6=B(\Xi_{c}(3080)^{+} \to \Lambda D^+)/B(\Xi_{c}(3080)^{+} \to \Sigma_{c}(2455)^{++} K^-)$ and large predicted decay width are not possible. Assignment with $\acute{\Xi}_{c1}^{'}(\frac{1}{2}^+)$ or $\acute{\Xi}_{c1}^{'}(\frac{3}{2}^+)$ is not possible for the large deviations of $\mathcal{B}_4=B(\Xi_{c}(3080)^{+} \to \Sigma_c(2520)^{++} K^-)/B(\Xi_{c}(3080)^{+} \to \Sigma_{c}(2455)^{++} K^-)$ and $\mathcal{B}_6=B(\Xi_{c}(3080)^{+} \to \Lambda D^+)/B(\Xi_{c}(3080)^{+} \to \Sigma_{c}(2455)^{++} K^-)$ between theory and experiment.

In comparison with experimental data, $\Xi_c(3080)^+$ is very possibly a $2S$-wave excitation $\acute{\Xi}_{c0}(\frac{1}{2}^+)$. In this assignment, the decay width $\Gamma=(3.6 \sim 31)$ MeV is compatible with the measurement, and the ratios $\mathcal{B}_4=B(\Xi_{c}(3080)^{+} \to \Sigma_c(2520)^{++} K^-)/B(\Xi_{c}(3080)^{+} \to \Sigma_{c}(2455)^{++} K^-)= 1.2 \sim 0.6$ and $\mathcal{B}_6=B(\Xi_{c}(3080)^{+} \to \Lambda D^+)/B(\Xi_{c}(3080)^{+} \to \Sigma_{c}(2455)^{++} K^-)= 1.7 \sim 3.1$ are in good agreement with the experimental measurements $1.07\pm0.27(stat)\pm0.04(sys)$ and $1.29\pm0.30(stat)\pm0.15(sys)$, respectively. In particular, the $\Xi_c\pi$ and $\Lambda_c K$ channels vanish in the assignment.

\begin{center}
  \begin{table*}\tiny
     \caption{Decay width (MeV) of $ \Xi_c(3080)^+$ as the $2S$-wave excitation in different assignments.
     $\mathcal{B}_4=B(\Xi_{c}(3080)^{+} \to \Sigma_c(2520)^{++} K^-)/B(\Xi_{c}(3080)^{+} \to \Sigma_{c}(2455)^{++} K^-)$;
     $\mathcal{B}_5=B(\Xi_{c}(3080)^{+} \to \Xi_c(2645)^{0} \pi^+)/B(\Xi_{c}(3080)^{+} \to \Sigma_{c}(2455)^{++} K^-)$;
     $\mathcal{B}_6=B(\Xi_{c}(3080)^{+} \to \Lambda D^+)/B(\Xi_{c}(3080)^{+} \to \Sigma_{c}(2455)^{++} K^-)$}
     \begin{tabular}{c |ccccc|cccccccccccccccccc} \hline \hline
 $ \Xi_{cJ_l} (J^P) $ & $ n_{\lambda} $ & $L_\lambda$  & $ n_{\rho} $ & $L_\rho$   & $S_\rho$
 &$\Xi_c\pi        $ &$\Xi_c^{'}\pi $  &$\Xi_c(2645)\pi$       &$\Sigma_c(2455) K$        &$\Sigma_c(2520) K $
 &$\Lambda_c K$      &$\Lambda D^+$    &$\Gamma_{sum}$     & $\mathcal{B}_4 $ &  $\mathcal{B}_5 $  &$\mathcal{B}_6  $   \\
     \hline

\label{01}$\acute{\Xi}_{c1}^{'}(\frac{1}{2}^+)$  & 1 &  0 &  0 & 0   &  1      &$0.3\sim3.4$   &$0.04\sim4.4$   &$0.1\sim1.7$        &$1.3\sim8.2$     &$0.4\sim1.2$      &$0.2\sim2.8$    &$0.3\sim4.3$    &$2.7\sim26$    &$0.3\sim0.1$    &$0.1\sim0.2$   &$0.4\sim0.8$\\  

\label{02} $\acute{\Xi}_{c1}^{'}(\frac{3}{2}^+)$  & 1 &  0 &  0 & 0  &  1      &$0.3\sim3.4$   &$0.01\sim1.1$   &$0.4\sim4.4$        &$0.3\sim2.1$     &$1.0\sim2.9$      &$0.2\sim2.8$    &$1.4\sim17$     &$3.5\sim34$    &$3.1\sim1.4$    &$1.2\sim2.1$   &$6.6\sim13$\\  

\label{03}$\acute{\Xi}_{c0}(\frac{1}{2}^+)$      & 1 &  0 &  0 & 0   &  0      &0              &$0.03\sim3.3$   &$0.4\sim5.2$        &$0.9\sim6.2$     &$1.2\sim3.5$      &0                &$1.0\sim13$    &$3.6\sim31$    &$1.2\sim0.6$    &$0.5\sim0.8$   &$1.7\sim3.1$    \\  

\label{04}$\tilde{\Xi}_{c1}(\frac{1}{2}^+)$& 0 &0 &  1 & 0   &  1      &$2.3\sim31$    &$0.4\sim39$     &$1.3\sim16$        &$11\sim74$       &$3.5\sim10$       &$1.9\sim25$      &0              &$22\sim196$    &$0.3\sim0.1$    &$0.1\sim0.2$     &$0$\\  

\label{05}$\tilde{\Xi}_{c1}(\frac{3}{2}^+)$& 0 &0 &  1 & 0   &  1      &$2.3\sim31$    &$0.1\sim9.9$    &$3.2\sim39$       &$2.8\sim19$      &$8.7\sim26$       &$1.9\sim25$      &0              &$19\sim150$    &$3.1\sim1.4$    &$1.2\sim2.1$   &0 \\  

\label{06}$\tilde{\Xi}_{c0}^{'}(\frac{1}{2}^+)$& 0&0&1 & 0   &  0      &0              &$0.3\sim30$    &$3.9\sim47$      &$8.5\sim56$      &$10\sim31$        &0                &0              &$23\sim163$    &$1.2\sim0.6$   &$0.5\sim0.8$   &0 \\  

    \hline\hline
    \end{tabular}
    \label{summary_simfit}
    \label{table8}
  \end{table*}
\end{center}

Numerical results of $\Xi_c(3080)$ as a $D$-wave excitation are given in Table~\ref{table9}. In most analyses of the $\Xi_c$ baryon spectrum, $\Xi_c(3080)$ was interpreted as a $D$-wave excitation of $\Xi_c$ with $J^P={5\over 2}^+$~\cite{ebert,cheng,guo,zhang1,zhang2,zhu2} or $J^P={5\over 2}^-$~\cite{zhu1}. In Ref.~\cite{zhong}, $\Xi_c(3080)^+$ was identified as the $\rho$-mode first radial excitation $(2S)$ of $\Xi_c$ in the chiral quark model. In Ref.~\cite{chen}, the theoretical result indicated that $\Xi_c(3080)^+$ could not be a $D$-wave $\Xi_c$ baryon with $J^P={5\over 2}^+$. From Table~\ref{table9}, none of the theoretical result agrees with experimental data, which implies that $\Xi_c(3080)^+$ can not be identified as a $D$-wave $\Xi_c$.

\begin{center}
  \begin{table*}\tiny
     \caption{Decay width (MeV) of $ \Xi_c(3080)^+$ as the $D$-wave excitation in different assignments. $\mathcal{B}_4=B(\Xi_{c}(3080)^{+} \to \Sigma_c(2520)^{++} K^-)/B(\Xi_{c}(3080)^{+} \to \Sigma_{c}(2455)^{++} K^-)$;
     $\mathcal{B}_5=B(\Xi_{c}(3080)^{+} \to \Xi_c(2645)^{0} \pi^+)/B(\Xi_{c}(3080)^{+} \to \Sigma_{c}(2455)^{++} K^-)$;
     $\mathcal{B}_6=B(\Xi_{c}(3080)^{+} \to \Lambda D^+)/B(\Xi_{c}(3080)^{+} \to \Sigma_{c}(2455)^{++} K^-)$}
     \begin{tabular}{c |ccccc|cccccccccccccccccc} \hline \hline
 $ \Xi_{cJ_l} (J^P) $ & $ n_{\lambda} $ & $L_\lambda$  & $ n_{\rho} $ & $L_\rho$   & $S_\rho$
 &$\Xi_c\pi        $ &$\Xi_c^{'}\pi $  &$\Xi_c(2645)\pi$       &$\Sigma_c(2455) K$        &$\Sigma_c(2520) K $
 &$\Lambda_c K$      &$\Lambda D^+$    &$\Gamma_{sum}$     & $\mathcal{B}_4 $ &  $\mathcal{B}_5 $  &$\mathcal{B}_6  $  \\
     \hline

\label{01}$\Xi_{c1}^{' }(\frac{1}{2}^+)$  & 0 &  2 &  0 & 0  &  1      &$0.7\sim7.8$   &$0.6\sim2.1$    &$0.3\sim0.8$        &$1.9\sim3.5$     &$0.6\sim0.7$     &$0.6\sim6.4$     &$20\sim56$      &$24\sim77$    &$0.3\sim0.2$    &$0.2\sim0.2$   &$15\sim24$    \\  

\label{02}$\Xi_{c1}^{' }(\frac{3}{2}^+)$  & 0 &  2 &  0 & 0  &  1      &$0.7\sim7.8$   &$0.2\sim0.5$    &$0.9\sim1.9$        &$0.5\sim0.9$     &$1.5\sim1.8$     &$0.6\sim6.4$     &$1.2\sim3.5$    &$5.6\sim23$    &$3.2\sim2.1$    &$1.8\sim2.2$   &$3.8\sim6.0$    \\  

\label{03}$\Xi_{c2}^{' }(\frac{3}{2}^+)$  & 0 &  2 &  0 & 0  &  1      &0              &$1.3\sim4.7$    &$0.7\sim0.8$        &$4.4\sim7.8$     &$0.6\sim0.7$      &0               &$11\sim31$      &$18\sim45$     &$0.1\sim0.1$    &$0.2\sim0.1$   &$3.8\sim6.0$    \\  

\label{04}$\Xi_{c2}^{' }(\frac{5}{2}^+)$  & 0 &  2 &  0 & 0  &  1      &0              &$0.5\sim0.2$    &$2.0\sim4.2$        &$0.3\sim0.08$    &$3.3\sim3.9$      &0                &$0.9\sim0.3$   &$7.0\sim8.6$   &$11\sim49$      &$6.5\sim51$     &$4.4\sim5.8$    \\  

\label{05}$\Xi_{c3}^{' }(\frac{5}{2}^+)$  & 0 &  2 &  0 & 0  &  1      &$2.6\sim1.2$   &$0.6\sim0.2$    &$0.3\sim0.08$       &$0.4\sim0.09$    &$0.04\sim0.01$    &$2.1\sim1.0$     &$0.07\sim0.02$ &$6.0\sim2.5$   &$0.1\sim0.1$    &$0.8\sim0.9$    &$0.3\sim0.4$    \\  

\label{06}$\Xi_{c3}^{' }(\frac{7}{2}^+)$  & 0 &  2 &  0 & 0  &  1      &$2.6\sim1.2$   &$0.3\sim0.1$    &$0.4\sim0.1$      &$0.2\sim0.05$    &$0.06\sim0.01$    &$2.1\sim1.0$     &$2.4\sim0.8$   &$8.0\sim3.2$   &$0.3\sim0.3$   &$2.0\sim2.2$     &$18\sim23$    \\  

\label{07}$\Xi_{c2}^{  }(\frac{3}{2}^+)$  & 0 &  2 &  0 & 0  &  0      &0              &$0.9\sim3.1$    &$0.8\sim0.6$        &$2.9\sim5.2$     &$0.5\sim0.5$      &0                &$7.3\sim21$    &$12\sim30$     &$0.2\sim0.1$    &$0.3\sim0.1$   &$3.8\sim6.0$ \\  

\label{08}$\Xi_{c2}^{  }(\frac{5}{2}^+)$  & 0 &  2 &  0 & 0  &  0      &0              &$0.7\sim0.3$    &$1.5\sim2.8$        &$0.5\sim0.1$     &$2.2\sim2.6$      &0                &$1.4\sim0.5$   &$6.3\sim6.2$   &$4.9\sim22$     &$3.2\sim23$    &$4.4\sim5.8$ \\  

\label{09}$\hat\Xi_{c1}^{' }(\frac{1}{2}^+)$ & 0 &  0 &  0 & 2  &  1   &$6.5\sim69$    &$5.3\sim18$     &$3.1\sim6.8$        &$17\sim31$       &$5.5\sim6.4$      &$5.3\sim56$      &0              &$43\sim188$    &$0.3\sim0.2$    &$0.2\sim0.2$  &0 \\  

\label{10}$\hat\Xi_{c1}^{' }(\frac{3}{2}^+)$ & 0 &  0 &  0 & 2  &  1   &$6.5\sim69$    &$1.3\sim4.6$    &$7.6\sim17$        &$4.3\sim7.7$     &$14\sim16$        &$5.3\sim56$      &0              &$39\sim171$    &$3.2\sim2.1$    &$1.8\sim2.2$  &0\\  

\label{11}$\hat\Xi_{c2}^{' }(\frac{3}{2}^+)$ & 0 &  0 &  0 & 2  &  1   &0              &$12\sim41$      &$6.1\sim7.1$        &$39\sim70$       &$5.4\sim5.9$       &0                &0             &$63\sim124$    &$0.1\sim0.1$    &$0.2\sim0.1$  &0\\  

\label{12}$\hat\Xi_{c2}^{' }(\frac{5}{2}^+)$ & 0 &  0 &  0 & 2  &  1   &0              &$4.5\sim1.6$    &$18\sim37$          &$2.8\sim0.8$     &$30\sim35$         &0               &0              &$55\sim74$     &$11\sim46$      &$6.4\sim48$  &0\\  

\label{13}$\hat\Xi_{c3}^{' }(\frac{5}{2}^+)$ & 0 &  0 &  0 & 2  &  1   &$23\sim11$     &$5.1\sim1.8$    &$2.7\sim0.8$       &$3.2\sim0.9$     &$0.4\sim0.09$     &$19\sim9.1$      &0              &$54\sim24$     &$0.1\sim0.1$    &$0.8\sim0.9$ &0\\  

\label{14}$\hat\Xi_{c3}^{' }(\frac{7}{2}^+)$ & 0 &  0 &  0 & 2  &  1   &$23\sim11$     &$2.9\sim1.0$    &$3.6\sim1.1$        &$1.8\sim0.5$     &$0.5\sim0.1$      &$19\sim9.1$      &0              &$51\sim23$     &$0.3\sim0.3$    &$2.0\sim2.2$ &0\\  

\label{15}$\hat\Xi_{c2}^{ }(\frac{3}{2}^+)$  & 0 &  0 &  0 & 2  &  0   &0              &$8.0\sim28$     &$6.9\sim5.6$        &$26\sim46$       &$4.0\sim4.0$      &0                &0              &$45\sim84$     &$0.2\sim0.1$    &$0.3\sim0.1$  &0\\  

\label{16}$\hat\Xi_{c2}^{ }(\frac{5}{2}^+)$  & 0 &  0 &  0 & 2  &  0   &0              &$6.7\sim2.4$    &$13\sim25$         &$4.2\sim1.1$    &$20\sim23$         &0                &0              &$44\sim52$     &$4.8\sim20$     &$3.1\sim22$   &0\\  

\label{17}$\check\Xi_{c0}^{'0}(\frac{1}{2}^+)$ & 0 &  1 &  0 & 1  &  0 &0              &$0.2\sim20$     &$2.6\sim31$         &$5.7\sim37$     &$12\sim33$         &0                &0              &$21\sim121$    &$2.2\sim0.9$    &$0.5\sim0.8$  &0\\  

\label{18}$\check\Xi_{c1}^{'1}(\frac{1}{2}^+)$ & 0 &  1 &  0 & 1  &  0 &0              &0             &0                 &0              &0       &0   &0       &0  &$\cdots$ &$\cdots$ &$\cdots$\\  
\label{19}$\check\Xi_{c1}^{'1}(\frac{3}{2}^+)$ & 0 &  1 &  0 & 1  &  0 &0              &0             &0                 &0             &0       &0   &0       &0  &$\cdots$ &$\cdots$ &$\cdots$\\  

\label{20}$\check\Xi_{c2}^{'2}(\frac{3}{2}^+)$ & 0 &  1 &  0 & 1  &  0 &0              &$5.3\sim18$     &$4.6\sim3.7$        &$17\sim31$       &$2.7\sim2.7$      &0                &0              &$30\sim56$     &$0.2\sim0.1$    &$0.3\sim0.1$  &0\\  

\label{21}$\check\Xi_{c2}^{'2}(\frac{5}{2}^+)$ & 0 &  1 &  0 & 1  &  0 &0              &$4.5\sim1.6$    &$8.8\sim17$        &$2.8\sim0.8$     &$13\sim15$        &0                &0              &$29\sim35$     &$4.8\sim20$     &$3.1\sim22$   &0\\  

\label{22}$\check\Xi_{c1}^{\ 0}(\frac{1}{2}^+)$& 0 &  1 &  0 & 1  &  1 &$1.6\sim21$    &$0.3\sim26$     &$0.9\sim10$        &$7.6\sim49$      &$4.0\sim11$       &$1.3\sim17$      &0              &$16\sim135$    &$0.5\sim0.2$    &$0.1\sim0.2$   &0\\  

\label{23}$\check\Xi_{c1}^{\ 0}(\frac{3}{2}^+)$& 0 &  1 &  0 & 1  &  1 &$1.6\sim21$    &$0.07\sim6.6$   &$2.1\sim26$         &$1.9\sim12$      &$10\sim27$        &$1.3\sim17$      &0              &$17\sim110$    &$5.4\sim2.2$    &$1.2\sim2.1$  &0\\  

\label{24}$\check\Xi_{c0}^{\ 1}(\frac{1}{2}^+)$ & 0 &  1 &  0 & 1  &  1&0              &0             &0        &0     &0     &0    &0       &0 &$\cdots$ &$\cdots$ &$\cdots$\\  

\label{25}$\check\Xi_{c1}^{\ 1}(\frac{1}{2}^+)$ & 0 &  1 &  0 & 1  &  1&0              &0             &0        &0     &0     &0    &0       &0 &$\cdots$ &$\cdots$ &$\cdots$\\  
\label{26}$\check\Xi_{c1}^{\ 1}(\frac{3}{2}^+)$ & 0 &  1 &  0 & 1  &  1&0              &0             &0        &0     &0     &0    &0       &0 &$\cdots$ &$\cdots$ &$\cdots$\\  
\label{27}$\check\Xi_{c2}^{\ 1}(\frac{3}{2}^+)$ & 0 &  1 &  0 & 1  &  1&0              &0             &0        &0     &0     &0    &0       &0 &$\cdots$ &$\cdots$ &$\cdots$\\  
\label{28}$\check\Xi_{c2}^{\ 1}(\frac{5}{2}^+)$ & 0 &  1 &  0 & 1  &  1&0              &0             &0        &0     &0     &0    &0       &0 &$\cdots$ &$\cdots$ &$\cdots$\\  

\label{29}$\check\Xi_{c1}^{\ 2}(\frac{1}{2}^+)$ & 0 &  1 &  0 & 1  &  1&$4.3\sim46$    &$3.5\sim12$     &$2.0\sim4.5$         &$12\sim21$       &$3.6\sim4.3$   &$3.5\sim37$     &0                  &$29\sim125$    &$0.3\sim0.2$    &$0.2\sim0.2$  &0\\  

\label{30}$\check\Xi_{c1}^{\ 2}(\frac{3}{2}^+)$ & 0 &  1 &  0 & 1  &  1&$4.3\sim46$    &$0.9\sim3.1$    &$5.1\sim11$         &$2.9\sim5.2$     &$9.1\sim11$    &$3.5\sim37$     &0                  &$26\sim114$    &$3.2\sim2.1$    &$1.8\sim2.2$  &0\\  

\label{31}$\check\Xi_{c2}^{\ 2}(\frac{3}{2}^+)$ & 0 &  1 &  0 & 1  &  1&0              &$8.0\sim28$     &$4.1\sim4.8$         &$26\sim46$       &$3.6\sim3.9$   &0               &0                  &$42\sim83$     &$0.1\sim0.1$    &$0.2\sim0.1$  &0\\  

\label{32}$\check\Xi_{c2}^{\ 2}(\frac{5}{2}^+)$ & 0 &  1 &  0 & 1  &  1&0              &$3.0\sim1.1$    &$12\sim25$          &$1.9\sim0.5$     &$20\sim23$      &0               &0                  &$37\sim49$    &$11\sim46$      &$6.4\sim48$  &0\\  

\label{33}$\check\Xi_{c3}^{\ 2}(\frac{5}{2}^+)$ & 0 &  1 &  0 & 1  &  1&$16\sim7.3$    &$3.4\sim1.2$    &$1.8\sim0.5$        &$2.1\sim0.6$     &$0.3\sim0.06$   &$13\sim4.6$    &0                  &$36\sim16$     &$0.1\sim0.1$    &$0.8\sim0.9$ &0\\  

\label{34}$\check\Xi_{c3}^{\ 2}(\frac{7}{2}^+)$ & 0 &  1 &  0 & 1  &  1&$16\sim7.3$    &$1.9\sim0.7$    &$2.4\sim0.7$        &$1.2\sim0.3$     &$0.4\sim0.08$   &$13\sim4.6$    &0                  &$34\sim15$     &$0.3\sim0.3$    &$2.0\sim2.2$    &0\\  
    \hline\hline
    \end{tabular}
    \label{summary_simfit}
    \label{table9}
  \end{table*}
\end{center}

\subsubsection{ $\Xi_c(3123)$}

$\Xi_c(3123)$ with a width of $4.4\pm 3.4\pm 1.7 $ MeV was observed by BaBar collaboration in the $ \Lambda_c^+ K^- \pi^+$ mass spectrum with a significance of $3.6$ standard deviations~\cite{babar1}, but no evidence for this state was found by Belle collaboration~\cite{belle3}. It decays through the intermediate resonant mode $\Sigma_c(2520)^{++} K^- $, and there was no evidence in the $\Lambda_c^+ K^-$ channel. Numerical results of $\Xi_c(3123)$ are given in Table~\ref{table10} and Table~\ref{table11}. For the poor experimental information of $\Xi_c(3123)$, it is impossible to identify this state. However, we go ahead with the following analyses under an assumption that $\Xi_c(3123)$ has a small total decay width (not larger than $10$ MeV).

\begin{center}
  \begin{table*}\tiny
     \caption{Decay width (MeV) of $ \Xi_c(3123)^+$ as the $2S$-wave excitation in different assignments. $\mathcal{B}_4=B(\Xi_{c}(3123)^{+} \to \Sigma_c(2520)^{++} K^-)/B(\Xi_{c}(3123)^{+} \to \Sigma_{c}(2455)^{++} K^-)$;
     $\mathcal{B}_5=B(\Xi_{c}(3123)^{+} \to \Xi_c(2645)^{0} \pi^+)/B(\Xi_{c}(3123)^{+} \to \Sigma_{c}(2455)^{++} K^-)$;
     $\mathcal{B}_6=B(\Xi_{c}(3123)^{+} \to \Lambda D^+)/B(\Xi_{c}(3123)^{+} \to \Sigma_{c}(2455)^{++} K^-)$}
     \begin{tabular}{c |ccccc|cccccccccccccccccc} \hline \hline
 $ \Xi_{cJ_l} (J^P) $ & $ n_{\lambda} $ & $L_\lambda$  & $ n_{\rho} $ & $L_\rho$   & $S_\rho$
 &$\Xi_c\pi        $ &$\Xi_c^{'}\pi $  &$\Xi_c(2645)\pi$       &$\Sigma_c(2455) K$        &$\Sigma_c(2520) K $
 &$\Lambda_c K$      &$\Lambda D^+$    &$\Gamma_{sum}$    & $\mathcal{B}_4 $ &  $\mathcal{B}_5 $  &$\mathcal{B}_6  $  \\
     \hline

\label{01}$\acute{\Xi}_{c1}^{'}(\frac{1}{2}^+)$  & 1 &  0 &  0 & 0   &  1      &$0.7\sim3.1$   &$0.01\sim4.7$   &$0.05\sim2.1$        &$0.5\sim11$      &$0.4\sim2.2$      &$0.7\sim2.4$     &$0.0\sim5.1$   &$2.4\sim30$    &$0.8\sim0.2$    &$0.1\sim0.2$   &$0.0\sim0.7$\\  

\label{02} $\acute{\Xi}_{c1}^{'}(\frac{3}{2}^+)$  & 1 &  0 &  0 & 0  &  1      &$0.7\sim3.1$   &$0.0\sim1.2$    &$0.1\sim5.2$        &$0.1\sim2.7$     &$1.0\sim5.5$      &$0.7\sim2.4$      &$0.01\sim20$  &$2.7\sim40$    &$7.6\sim2.1$    &$1.1\sim1.9$   &$0.1\sim11$\\  

\label{03}$\acute{\Xi}_{c0}(\frac{1}{2}^+)$      & 1 &  0 &  0 & 0   &  0      &0              &$0.01\sim3.5$   &$0.2\sim6.2$        &$0.4\sim8.0$     &$1.2\sim6.7$      &0                &$0.01\sim15$   &$1.8\sim40$    &$3.0\sim0.8$    &$0.4\sim0.8$   &$0.0\sim2.8$    \\  

\label{04}$\tilde{\Xi}_{c1}(\frac{1}{2}^+)$& 0 &0 &  1 & 0   &  1      &$6.6\sim28$    &$0.08\sim42$     &$0.5\sim19$        &$4.9\sim96$      &$3.6\sim20$       &$6.3\sim22$      &0              &$22\sim227$    &$0.8\sim0.2$     &$0.1\sim0.2$     &$0$\\  

\label{05}$\tilde{\Xi}_{c1}(\frac{3}{2}^+)$& 0 &0 &  1 & 0   &  1      &$6.6\sim28$    &$0.02\sim11$     &$1.2\sim47$       &$1.2\sim24$      &$9.1\sim50$       &$6.3\sim22$      &0              &$24\sim181$    &$7.6\sim2.1$     &$1.1\sim1.9$   &0 \\  

\label{06}$\tilde{\Xi}_{c0}^{'}(\frac{1}{2}^+)$& 0&0&1 & 0   &  0      &0              &$0.06\sim31$    &$1.5\sim56$      &$3.7\sim72$      &$11\sim60$        &0                &0              &$16\sim220$    &$3.0\sim0.8$    &$0.4\sim0.8$   &0 \\  

    \hline\hline
    \end{tabular}
    \label{summary_simfit}
    \label{table10}
  \end{table*}
\end{center}

From Table~\ref{table10}, $\Xi_c(3123)$ seems impossible a $\acute{\Xi}_{c0}(\frac{1}{2}^+)$, $\tilde{\Xi}_{c1}(\frac{3}{2}^+)$ or $\tilde{\Xi}_{c0}^{'}(\frac{1}{2}^+)$ for a large predicted decay width. For a small decay width, $\Xi_c(3123)$ may be a $\acute{\Xi}_{c1}^{'}(\frac{1}{2}^+)$, $\acute{\Xi}_{c1}^{'}(\frac{3}{2}^+)$ or $\acute{\Xi}_{c0}(\frac{1}{2}^+)$. In these assignments, the observation of $\Xi_c\pi$ and $\Lambda_c K$ modes and measurements of the branching fraction ratios $\mathcal{B}_4$, $\mathcal{B}_5$ and $\mathcal{B}_6$ are important to identify this state.

\begin{center}
  \begin{table*}\tiny
     \caption{Decay width (MeV) of $ \Xi_c(3123)^+$ as the $D$-wave excitation in different assignments. $\mathcal{B}_4=B(\Xi_{c}(3123)^{+} \to \Sigma_c(2520)^{++} K^-)/B(\Xi_{c}(3123)^{+} \to \Sigma_{c}(2455)^{++} K^-)$;
     $\mathcal{B}_5=B(\Xi_{c}(3123)^{+} \to \Xi_c(2645)^{0} \pi^+)/B(\Xi_{c}(3123)^{+} \to \Sigma_{c}(2455)^{++} K^-)$;
     $\mathcal{B}_6=B(\Xi_{c}(3123)^{+} \to \Lambda D^+)/B(\Xi_{c}(3123)^{+} \to \Sigma_{c}(2455)^{++} K^-)$}
     \begin{tabular}{c |ccccc|cccccccccccccccccc} \hline \hline
 $ \Xi_{cJ_l} (J^P) $ & $ n_{\lambda} $ & $L_\lambda$  & $ n_{\rho} $ & $L_\rho$   & $S_\rho$
 &$\Xi_c\pi        $ &$\Xi_c^{'}\pi $  &$\Xi_c(2645)\pi$       &$\Sigma_c(2455) K$        &$\Sigma_c(2520) K $
 &$\Lambda_c K$      &$\Lambda D^+$    &$\Gamma_{sum}$     & $\mathcal{B}_4 $ &  $\mathcal{B}_5 $  &$\mathcal{B}_6  $   \\
     \hline

\label{01}$\Xi_{c1}^{' }(\frac{1}{2}^+)$  & 0 &  2 &  0 & 0  &  1      &$0.3\sim8.0$   &$0.5\sim2.4$    &$0.3\sim1.0$        &$1.9\sim4.8$     &$0.9\sim1.5$     &$0.2\sim6.4$     &$16\sim78$      &$20\sim102$    &$0.5\sim0.3$    &$0.2\sim0.2$   &$13\sim24$    \\  

\label{02}$\Xi_{c1}^{' }(\frac{3}{2}^+)$  & 0 &  2 &  0 & 0  &  1      &$0.3\sim8.0$   &$0.1\sim0.6$    &$0.8\sim2.4$        &$0.5\sim1.2$     &$2.3\sim3.6$     &$0.2\sim6.4$     &$1.0\sim4.9$    &$5.2\sim27$    &$5.0\sim3.0$    &$1.7\sim2.0$   &$3.2\sim6.1$    \\  

\label{03}$\Xi_{c2}^{' }(\frac{3}{2}^+)$  & 0 &  2 &  0 & 0  &  1      &0              &$1.0\sim5.3$    &$1.0\sim1.1$        &$4.2\sim11$      &$1.2\sim1.4$      &0                &$8.8\sim44$    &$16\sim62$     &$0.3\sim0.1$    &$0.2\sim0.1$   &$3.2\sim6.1$    \\  

\label{04}$\Xi_{c2}^{' }(\frac{5}{2}^+)$  & 0 &  2 &  0 & 0  &  1      &0              &$0.8\sim0.3$    &$2.0\sim5.3$        &$0.8\sim0.2$     &$5.1\sim7.9$      &0                &$2.7\sim1.2$   &$11\sim15$     &$6.4\sim33$     &$2.5\sim22$     &$5.1\sim7.1$    \\  

\label{05}$\Xi_{c3}^{' }(\frac{5}{2}^+)$  & 0 &  2 &  0 & 0  &  1      &$3.5\sim1.8$   &$0.9\sim0.4$    &$0.6\sim0.2$       &$0.9\sim0.3$     &$0.3\sim 0.07$    &$3.0\sim1.6$     &$0.2\sim0.08$  &$9.3\sim4.4$   &$0.3\sim0.2$    &$0.6\sim0.6$     &$0.3\sim0.4$    \\  

\label{06}$\Xi_{c3}^{' }(\frac{7}{2}^+)$  & 0 &  2 &  0 & 0  &  1      &$3.5\sim1.8$   &$0.5\sim0.2$    &$0.8\sim0.2$      &$0.5\sim0.2$     &$0.4\sim0.09$     &$3.0\sim1.6$     &$7.0\sim3.0$   &$16\sim7.1$    &$0.7\sim0.6$    &$1.4\sim1.5$    &$20\sim29$    \\  

\label{07}$\Xi_{c2}^{  }(\frac{3}{2}^+)$  & 0 &  2 &  0 & 0  &  0      &0              &$0.7\sim3.6$    &$1.2\sim0.9$        &$2.8\sim7.2$     &$1.1\sim1.0$      &0                &$5.9\sim29$    &$12\sim42$     &$0.4\sim0.1$    &$0.4\sim0.1$   &$3.2\sim6.1$ \\  

\label{08}$\Xi_{c2}^{  }(\frac{5}{2}^+)$  & 0 &  2 &  0 & 0  &  0      &0              &$1.2\sim0.5$    &$1.6\sim3.6$        &$0.2\sim0.4$     &$3.5\sim5.3$      &0                &$4.1\sim1.7$   &$12\sim11$     &$3.0\sim15$     &$1.3\sim9.8$    &$5.1\sim7.1$ \\  

\label{09}$\hat\Xi_{c1}^{' }(\frac{1}{2}^+)$ & 0 &  0 &  0 & 2  &  1   &$2.8\sim70$    &$4.1\sim21$     &$2.9\sim8.5$        &$17\sim43$       &$8.3\sim13$       &$1.7\sim55$      &0              &$36\sim210$    &$0.5\sim0.3$    &$0.2\sim0.2$  &0 \\  

\label{10}$\hat\Xi_{c1}^{' }(\frac{3}{2}^+)$ & 0 &  0 &  0 & 2  &  1   &$2.8\sim70$    &$1.0\sim5.2$    &$7.1\sim21$        &$4.2\sim11$      &$21\sim32$        &$1.7\sim55$      &0              &$37\sim195$    &$5.0\sim3.1$    &$1.7\sim2.0$  &0\\  

\label{11}$\hat\Xi_{c2}^{' }(\frac{3}{2}^+)$ & 0 &  0 &  0 & 2  &  1   &0              &$9.2\sim47$      &$8.9\sim9.8$        &$37\sim96$       &$10\sim12$        &0                &0              &$66\sim165$    &$0.3\sim0.1$     &$0.2\sim0.1$  &0\\  

\label{12}$\hat\Xi_{c2}^{' }(\frac{5}{2}^+)$ & 0 &  0 &  0 & 2  &  1   &0              &$7.3\sim2.9$    &$18\sim47$          &$7.2\sim2.3$     &$46\sim70$         &0               &0              &$79\sim122$    &$6.3\sim31$     &$2.5\sim20$  &0\\  

\label{13}$\hat\Xi_{c3}^{' }(\frac{5}{2}^+)$ & 0 &  0 &  0 & 2  &  1   &$32\sim17$     &$8.3\sim3.3$    &$5.0\sim1.7$       &$8.3\sim2.6$     &$2.4\sim0.6$       &$27\sim15$       &0             &$83\sim40$     &$0.3\sim0.2$    &$0.6\sim0.6$ &0\\  

\label{14}$\hat\Xi_{c3}^{' }(\frac{7}{2}^+)$ & 0 &  0 &  0 & 2  &  1   &$32\sim17$     &$4.7\sim1.9$    &$6.8\sim2.3$        &$4.7\sim1.5$     &$3.3\sim0.8$       &$27\sim15$       &0             &$78\sim39$     &$0.7\sim0.6$    &$1.4\sim1.5$ &0\\  

\label{15}$\hat\Xi_{c2}^{ }(\frac{3}{2}^+)$  & 0 &  0 &  0 & 2  &  0   &0              &$6.1\sim31$     &$11\sim8.3$        &$25\sim64$       &$9.5\sim9.0$       &0                &0             &$52\sim113$    &$0.4\sim0.1$    &$0.4\sim0.1$  &0\\  

\label{16}$\hat\Xi_{c2}^{ }(\frac{5}{2}^+)$  & 0 &  0 &  0 & 2  &  0   &0              &$11\sim4.3$     &$15\sim32$         &$11\sim3.4$      &$32\sim47$         &0                &0             &$68\sim87$     &$2.9\sim14$     &$1.3\sim9.2$   &0\\  

\label{17}$\check\Xi_{c0}^{'0}(\frac{1}{2}^+)$ & 0 &  1 &  0 & 1  &  0 &0              &$0.04\sim21$    &$1.0\sim37$         &$2.4\sim48$     &$13\sim63$          &0                &0             &$17\sim170$    &$5.4\sim1.3$    &$0.4\sim0.8$  &0\\  

\label{18}$\check\Xi_{c1}^{'1}(\frac{1}{2}^+)$ & 0 &  1 &  0 & 1  &  0 &0              &0               &0                   &0                &0       &0   &0       &0   &$\cdots$ &$\cdots$ &$\cdots$ \\  
\label{19}$\check\Xi_{c1}^{'1}(\frac{3}{2}^+)$ & 0 &  1 &  0 & 1  &  0 &0              &0               &0                   &0                &0       &0   &0       &0   &$\cdots$ &$\cdots$ &$\cdots$ \\  

\label{20}$\check\Xi_{c2}^{'2}(\frac{3}{2}^+)$ & 0 &  1 &  0 & 1  &  0 &0              &$4.1\sim21$     &$7.5\sim5.5$        &$17\sim43$       &$6.4\sim6.0$     &0                &0               &$35\sim75$     &$0.4\sim0.1$    &$0.4\sim0.1$  &0\\  

\label{21}$\check\Xi_{c2}^{'2}(\frac{5}{2}^+)$ & 0 &  1 &  0 & 1  &  0 &0              &$7.3\sim2.9$    &$9.7\sim21$        &$7.2\sim2.3$     &$21\sim31$       &0                &0               &$45\sim58$     &$2.9\sim14$     &$1.3\sim9.2$   &0\\  

\label{22}$\check\Xi_{c1}^{\ 0}(\frac{1}{2}^+)$& 0 &  1 &  0 & 1  &  1 &$4.4\sim19$    &$0.06\sim28$    &$0.3\sim12$        &$3.3\sim64$      &$4.4\sim21$      &$4.2\sim14$      &0               &$17\sim159$    &$1.4\sim0.3$    &$0.1\sim0.2$   &0\\  

\label{23}$\check\Xi_{c1}^{\ 0}(\frac{3}{2}^+)$& 0 &  1 &  0 & 1  &  1 &$4.4\sim19$    &$0.01\sim7.0$   &$0.8\sim31$         &$0.8\sim16$      &$11\sim53$       &$4.2\sim14$      &0               &$21\sim140$    &$14\sim3.3$     &$1.1\sim1.9$  &0\\  

\label{24}$\check\Xi_{c0}^{\ 1}(\frac{1}{2}^+)$ & 0 &  1 &  0 & 1  &  1&0              &0               &0          &0       &0       &0    &0       &0    &$\cdots$ &$\cdots$ &$\cdots$\\  

\label{25}$\check\Xi_{c1}^{\ 1}(\frac{1}{2}^+)$ & 0 &  1 &  0 & 1  &  1&0              &0               &0          &0        &0       &0    &0       &0    &$\cdots$ &$\cdots$ &$\cdots$\\  
\label{26}$\check\Xi_{c1}^{\ 1}(\frac{3}{2}^+)$ & 0 &  1 &  0 & 1  &  1&0              &0               &0          &0       &0       &0    &0       &0    &$\cdots$ &$\cdots$ &$\cdots$\\  
\label{27}$\check\Xi_{c2}^{\ 1}(\frac{3}{2}^+)$ & 0 &  1 &  0 & 1  &  1&0              &0               &0          &0       &0       &0    &0       &0    &$\cdots$ &$\cdots$ &$\cdots$\\  
\label{28}$\check\Xi_{c2}^{\ 1}(\frac{5}{2}^+)$ & 0 &  1 &  0 & 1  &  1&0              &0               &0          &0       &0       &0    &0       &0    &$\cdots$ &$\cdots$ &$\cdots$\\  

\label{29}$\check\Xi_{c1}^{\ 2}(\frac{1}{2}^+)$ & 0 &  1 &  0 & 1  &  1&$1.9\sim47$    &$2.7\sim13$     &$1.9\sim5.7$         &$11\sim28$       &$5.5\sim8.6$    &$1.1\sim37$     &0                 &$24\sim140$    &$0.5\sim0.3$    &$0.2\sim0.2$  &0\\  

\label{30}$\check\Xi_{c1}^{\ 2}(\frac{3}{2}^+)$ & 0 &  1 &  0 & 1  &  1&$1.9\sim47$    &$0.7\sim3.5$    &$4.8\sim14$         &$2.8\sim7.1$     &$14\sim22$      &$1.1\sim37$     &0                 &$25\sim130$    &$5.0\sim3.1$    &$1.7\sim2.0$  &0\\  

\label{31}$\check\Xi_{c2}^{\ 2}(\frac{3}{2}^+)$ & 0 &  1 &  0 & 1  &  1&0              &$6.1\sim31$     &$5.9\sim6.5$         &$25\sim64$       &$7.0\sim8.3$    &0               &0                 &$44\sim110$    &$0.3\sim0.1$    &$0.2\sim0.1$  &0\\  

\label{32}$\check\Xi_{c2}^{\ 2}(\frac{5}{2}^+)$ & 0 &  1 &  0 & 1  &  1&0              &$4.8\sim1.9$    &$12\sim31$          &$4.8\sim1.5$     &$31\sim47$      &0               &0                 &$53\sim82$     &$6.3\sim31$     &$2.5\sim20$  &0\\  

\label{33}$\check\Xi_{c3}^{\ 2}(\frac{5}{2}^+)$ & 0 &  1 &  0 & 1  &  1&$21\sim11$     &$5.5\sim2.2$    &$3.4\sim1.1$        &$5.5\sim1.7$     &$1.6\sim0.4$    &$18\sim10$      &0                 &$55\sim27$     &$0.3\sim0.2$    &$0.6\sim0.6$ &0\\  

\label{34}$\check\Xi_{c3}^{\ 2}(\frac{7}{2}^+)$ & 0 &  1 &  0 & 1  &  1&$21\sim11$     &$3.1\sim1.2$    &$4.5\sim1.5$        &$3.1\sim1.0$     &$2.2\sim0.6$    &$18\sim10$      &0                 &$52\sim26$     &$0.7\sim0.6$    &$1.4\sim1.5$    &0\\  
    \hline\hline
    \end{tabular}
    \label{summary_simfit}
    \label{table11}
  \end{table*}
\end{center}

Numerical results in Table~\ref{table11} suggest that $\Xi_c(3123)$ is very possibly a $\Xi_{c1}^{' }(\frac{3}{2}^+)$ or $\Xi_{c2}^{  }(\frac{3}{2}^+)$ with quantum numbers $J^P={3\over 2}^+$. $\Xi_c(3123)$ could be a $\Xi_{c2}^{' }(\frac{5}{2}^+)$, $\Xi_{c3}^{' }(\frac{5}{2}^+)$ or $\Xi_{c2}^{  }(\frac{5}{2}^+)$ with quantum numbers $J^P={5\over 2}^+$. It may be a $\Xi_{c3}^{' }(\frac{7}{2}^+)$ with quantum numbers $J^P={7\over 2}^+$. In these assignments, the observation of $\Xi_c\pi$ and $\Lambda_c K$ modes and measurement of some branching fraction ratios are also important for its identification.

In the calculation of $P$-wave, $D$-wave and $2S$-wave of excited $\Xi_c$, it is found that only the excited $\Xi_c$ without $\rho$ mode excitation (with quantum numbers $(n_\rho,L_ \rho )=(0,0)$) may decay into $\Lambda D^+$. The width of other excited $\Xi_c$ with non-vanish $(n_\rho,L_ \rho )$ decaying into $\Lambda D^+$ mode vanishes.

\section{Conclusions and discussions}

Many excited charmed strange baryons $\Xi_c(2790)$, $\Xi_c(2815)$, $\Xi_c(2930)$, $\Xi_c(2980)$, $\Xi_c(3055)$, $\Xi_c(3080)$ and $\Xi_c(3123)$ have been reported and there are many interpretations of them. The strong decay properties of these baryons as possible $P$-wave excited $\Xi_c$ candidates have been systematically studied in a $^3P_0$ model in Ref.~\cite{ye1}. In the reference, $\Xi_c(2790)$ and $\Xi_c(2815)$ were assigned with a $P$-wave excited $\Xi_{c1}({1\over 2}^-)$ and $\Xi_{c1}({3\over 2}^-)$, respectively. $\Xi_c(2930)$ and $\Xi_c(2980)$ may be a $P$-wave excitation of $\Xi_c$, while $\Xi_c(3055)$, $\Xi_c(3080)$ and $\Xi_c(3123)$ can not be a $P$-wave excitation of $\Xi_c$. As an extension, the strong decay properties of $\Xi_c(2930)$, $\Xi_c(2980)$, $\Xi_c(3055)$, $\Xi_c(3080)$ and $\Xi_c(3123)$ as possible first radially excited $2S$-wave and $1D$-wave $\Xi_c$ baryons have been systematically studied in the $^3P_0$ model in this paper. In comparison with experimental data, possible configurations of $\Xi_c(2930)$, $\Xi_c(2980)$, $\Xi_c(3055)$, $\Xi_c(3080)$ and $\Xi_c(3123)$ have been analyzed. Some useful branching fraction ratios to distinguish these different configurations have also been calculated.

For $\Xi_c(2930)$, $\Xi_c(2980)$, $\Xi_c(3055)$, $\Xi_c(3080)$ and $\Xi_c(3123)$, the main conclusions are as follows:

$1$, $\Xi_c(2930)$ may be a $P$-wave, $D$-wave or $2S$-wave excitation of $\Xi_c$ though this signal was found only by BaBar collaboration.
From limited experimental information, $\Xi_c(2930)$ may be a $2S$-wave excitation, $\tilde{\Xi}_{c1}(\frac{1}{2}^+)$ or $\tilde{\Xi}_{c1}(\frac{3}{2}^+)$. In order to distinguish these two possibilities, measurements of the different ratios, $\mathcal{B}_2=B(\Xi_{c}(2930)^{+} \to \Xi_c^{'0} \pi^{+})/B(\Xi_{c}(2930)^{+} \to \Xi_{c}(2645)^{0}\pi^{+})=2.5\sim 4.1$ for $\tilde{\Xi}_{c1}(\frac{1}{2}^+)$ and the $\mathcal{B}_2=0.3\sim 0.4$ for $\tilde{\Xi}_{c1}(\frac{3}{2}^+)$, will be helpful.

It is also possibly a $D$-wave excited $\hat\Xi_{c1}^{' }(\frac{1}{2}^+)$, $\check\Xi_{c1}^{\ 0}(\frac{1}{2}^+)$ , $\check\Xi_{c1}^{\ 2}(\frac{1}{2}^+)$, $\hat\Xi_{c1}^{' }(\frac{3}{2}^+)$, $\check\Xi_{c1}^{\ 0}(\frac{3}{2}^+)$ or $\check\Xi_{c1}^{\ 2}(\frac{3}{2}^+)$. With these two different $J^P$ quantum numbers, the ratios $\mathcal{B}_2=B(\Xi_{c}(2930)^{+} \to \Xi_c^{'0} \pi^{+})/B(\Xi_{c}(2930)^{+} \to \Xi_{c}(2645)^{0}\pi^{+})$ are different. Therefore, measurements of these ratios in the future are important for the identification of $\Xi_c(2930)$.

$2$, $\Xi_c(2980)^+$ may be a $P$-wave, $D$-wave or $2S$-wave excitation of $\Xi_c$. $\Xi_c(2980)^+$ is very possibly a radial excitation $\tilde{\Xi}_{c1}(\frac{1}{2}^+)$ or $\tilde{\Xi}_{c0}^{'}(\frac{1}{2}^+)$ with $J^P={1\over 2}^+$. In these assignment, $ \Xi_c(2980)^+$ has a total decay width and branching fraction ratio $B'_2$ comparable with experimental data from Belle collaboration. $\Xi_c(2980)^+$ could be assigned as $\check\Xi_{c0}^{'0}(\frac{1}{2}^+)$ or $\check\Xi_{c1}^{\ 0}(\frac{1}{2}^+)$ with $J^P={1\over 2}^+$ in the case as a $D$-wave excitation. In these assignments, $ \Xi_c(2980)^+$ has a total decay width and branching fraction ratio $B'_2$ comparable with experimental data from the Belle collaboration.

$3$, $\Xi_c(3055)^+$ may be a $2S$-wave excitation $\acute{\Xi}_{c1}^{'}(\frac{3}{2}^+)$ or $\acute{\Xi}_{c0}(\frac{1}{2}^+)$. $\Xi_c(3055)^+$ can also be a $D$-wave excitation $\Xi_{c1}^{' }(\frac{3}{2}^+)$, $\Xi_{c2}^{' }(\frac{5}{2}^+)$, $\Xi_{c2}^{  }(\frac{3}{2}^+)$ or $\Xi_{c2}^{  }(\frac{5}{2}^+)$.

$4$, $\Xi_c(3080)^+$ is very possibly a $2S$-wave excitation, $\acute{\Xi}_{c0}(\frac{1}{2}^+)$, and seems not a $D$-wave excitation of $\Xi_c$. In this assignment, the decay width $\Gamma=(3.6 \sim 31)$ MeV is compatible with the measurement, and the ratios $\mathcal{B}_4=B(\Xi_{c}(3080)^{+} \to \Sigma_c(2520)^{++} K^-)/B(\Xi_{c}(3080)^{+} \to \Sigma_{c}(2455)^{++} K^-)= 1.2 \sim 0.6$ and $\mathcal{B}_6=B(\Xi_{c}(3080)^{+} \to \Lambda D^+)/B(\Xi_{c}(3080)^{+} \to \Sigma_{c}(2455)^{++} K^-)= 1.7 \sim 3.1$ are in good agreement with the experimental measurements $1.07\pm0.27(stat)\pm0.04(sys)$ and $1.29\pm0.30(stat)\pm0.15(sys)$, respectively. In particular, the $\Xi_c\pi$ and $\Lambda_c K$ channels vanish in this assignment.

$5$, for the poor experimental information of $\Xi_c(3123)$, it is not possible to identify this state. However, if $\Xi_c(3123)$ has a small decay width, it may be a $2S$-wave or $D$-wave excitation of $\Xi_c$. $\Xi_c(3123)$ could be a $2S$-wave excitation $\acute{\Xi}_{c1}^{'}(\frac{1}{2}^+)$, $\acute{\Xi}_{c1}^{'}(\frac{3}{2}^+)$ or $\acute{\Xi}_{c0}(\frac{1}{2}^+)$. It could be a $\Xi_{c1}^{' }(\frac{3}{2}^+)$ or $\Xi_{c2}^{  }(\frac{3}{2}^+)$ with $J^P={3\over 2}^+$. $\Xi_c(3123)$ could be a $\Xi_{c2}^{' }(\frac{5}{2}^+)$, $\Xi_{c3}^{' }(\frac{5}{2}^+)$ or $\Xi_{c2}^{  }(\frac{5}{2}^+)$ with $J^P={5\over 2}^+$. It may be a $\Xi_{c3}^{' }(\frac{7}{2}^+)$ with $J^P={7\over 2}^+$. In these assignments, the observation of $\Xi_c\pi$ and $\Lambda_c K$ modes and measurements of some branching fraction ratios are also important for its identification.

In order to identify these excited $\Xi_c$, more experiments on their confirmation, masses and total decay widths are required. In particular, some important branching fraction ratios for the identification must be measured also in the future.

\begin{acknowledgments}
 This work is supported by National Natural Science Foundation of China under the grants No: 11475111 and No: 11075102.
\end{acknowledgments}

\end{document}